\documentclass[journal]{IEEEtran}
\usepackage[style=ieee,maxnames=4,minnames=3,maxbibnames=3]{biblatex}

\usepackage{setspace}
\usepackage{amsfonts}
\usepackage[cmex10]{amsmath}
\usepackage{amsthm}

\usepackage{diagbox}   
\usepackage{multirow}
\usepackage{array}
\theoremstyle{definition}

\theoremstyle{remark}

\theoremstyle{remark}

\usepackage{pifont} 
\usepackage{array}
\usepackage{mdwmath}
\usepackage{mdwtab}
\usepackage{eqparbox}
\usepackage[caption=1]{caption}
\usepackage{subcaption}
\usepackage{url}
\usepackage{amssymb}
\usepackage{float}
\usepackage{indentfirst}
\usepackage{makeidx}
\usepackage{tabularx}
\usepackage{cases}
\usepackage{algorithmic}
\usepackage{graphicx}
\usepackage{subcaption} 
\usepackage{mathtools}
\usepackage{color}
\usepackage{bm}
\usepackage{stfloats}
\usepackage{lipsum}
\usepackage{amsfonts}
\usepackage{bm}
\usepackage{dsfont}
\usepackage{lipsum}
\usepackage{graphicx}
\usepackage{url}
\usepackage{bbm}
\usepackage{booktabs}
\usepackage{multirow} 
\usepackage{makecell} 
\newcommand{\RNum}[1]{\uppercase\expandafter{\romannumeral #1\relax}}
\ifCLASSOPTIONcompsoc
\usepackage[caption=false, font=normalsize, labelfont=sf, textfont=sf]{subfig}
\else
\usepackage[caption=false, font=footnotesize]{subfig}
\fi
\usepackage{multirow}
\allowdisplaybreaks
\makeatletter

\usepackage[linesnumbered,ruled,vlined]{algorithm2e}
\usepackage{amsmath}
\makeatother

\usepackage{amsthm}
\theoremstyle{definition} 
\theoremstyle{remark} 

\begin{document}
\title{Diffusion Inpainting MIMO-OFDM Channels with Limited Noisy Observations}
\author{
    Weijie~Zhou,~
    Zhaoyang~Zhang,~
    Yuzhi~Yang,~
    Sen~Yan,~
    Zhixian~Kong,~
    and~M\'erouane~Debbah
\thanks{W. Zhou, Z. Zhang (\textit{Corresponding Author}), and Z. Kong are with 1) College of Information Science and Electronic Engineering, Zhejiang University, Hangzhou 310027, China, and 2) Zhejiang Key Laboratory of Multi-modal Commu. Netw. \& Intell. Info. Proc., Hangzhou 310027, China (e-mails: \{wj\_zhou, ning\_ming, zhixian\_kong\}@zju.edu.cn). }
\thanks{Y. Yang, S. Yan, and M. Debbah are with College of Computing and Mathematical Sciences, Khalifa University, Abu Dhabi 127788, UAE (e-mails: \{yuzhi.yang, merouane.debbah\}@ku.ac.ae, yansen0508@gmail.com). }
}
\maketitle
\begin{abstract}
Acquiring the channel state information from limited and noisy observations at pilot positions is critical for wireless multiple-input multiple-output (MIMO)-orthogonal frequency division multiplexing (OFDM) systems. In this paper, we view this process as a conditional generative task in which the partial noisy channel estimates at the pilots are utilized as a ``prompt'' to guide the diffusion ``inpainting'' of the underlying channel.
To this end, we resort to a general Conditional Diffusion Transformer (CDiT) framework with a well-designed network architecture and update rule. 
In particular, we design a dedicated embedding strategy to encode and adapt to different pilot patterns and noise levels, and utilize a special cross-attention mechanism to align the partial raw channel observations with the denoised channel at each time step of the generation process.  
This architecture effectively anchors the diffusion process, enabling the model to accurately recover full channel details from limited noisy observations. 
Comprehensive experimental results show that, the proposed approach achieves a performance gain of over 5 dB compared to the baselines under varying noise conditions, and provides robust channel acquisition even under a sparse pilot density of 1/32 without significant performance loss compared to the denser pilot cases. Moreover, it is capable of generating high-quality channel matrices within just 10 inference steps, effectively balancing estimation accuracy with computational efficiency and inference speed. Ablation studies demonstrate the rationality of the model design and the necessity of its modules.
\end{abstract}

\section{Introduction}

\subsection{Background and Motivations}

Channel estimation is one of the fundamental problems in the multiple-input multiple-output (MIMO) - orthogonal frequency division multiplexing (OFDM) receiver. While linear estimation methods such as Least-Square (LS) and linear minimum mean-squared error (LMMSE) are widely used, they exhibit significant limitations. The former suffers from substantial performance degradation when the number of pilots decreases, while the latter requires the channel covariance matrix as prior information and has considerable computational overhead due to matrix inversion especially when dealing with high-dimensional channel data.

With the advancement of artificial intelligence (AI), neural networks (NNs) offer new approaches to address the channel estimation problem. Data-driven deep learning (DL) has been widely applied to tackle these problems \cite{10445518}.
However, in the channel estimation problem, the inputs are typically contaminated by noise, which may significantly degrade the quality of the input data \cite{yang2025generativediffusionreceiversachieving}.
Without specialized network design and training, noisy inputs usually lead to substantial performance degradation, especially when the noise power varies in a large range. Furthermore, when there are multiple data streams during the communication process, different streams use distinct carriers for pilot symbols. Therefore, NNs must also adapt to various pilot patterns, avoiding the need to train separate network parameters for each configuration. This adaptability represents one of the key aspects of applying NNs to channel estimation.

Diffusion models (DMs) have achieved tremendous success in the field of image processing, demonstrating excellent performance in tasks such as super-resolution, inpainting, denoising, inverse problem solving, text-to-image synthesis, etc. Enabled by the training procedure based on the diffusion process, diffusion models always have extraordinary noise tolerance capability with different noise levels.
Recently, a growing number of studies have focused on introducing DMs to inherent wireless communication problems such as channel estimation and data detection \cite{yang2025generativediffusionreceiversachieving, cai2025jointactivitydetectionchannel,10446413,9957135,10930691,10946972}. 
Applying DMs to these problems typically involves generating targets conditioned on incomplete channel estimates or received signals. However, they cannot be directly mapped to any steps within the standard diffusion process. This discrepancy necessitates additional modifications to the training strategies or generation processes to properly incorporate such guidance.

On the other hand, existing DM works in other tasks indicate that DMs possess significant potential for compatibility with diverse downstream tasks by integrating task information as conditions, which has not been sufficiently discussed in wireless receiver tasks. 
Through appropriate conditioning, DMs can take in information from various domains and show great adaptability in different scenarios.
In this paper, we refer from the DM-based image inpainting tasks, and reconsider the MIMO-OFDM channel estimation task as a prompt-guided generative task. 
In particular, we incorporate the parameters of communication systems, such as pilot patterns and noise levels, and raw estimates as conditional information in the DM-based receiver design to better guide generation and ensure adaptability.

When applying DMs to conditional generation problems, two primary approaches are typically employed. The first method involves preserving the original training process of the denoising network while adjusting the generation procedure \cite{9957135,10930691,10946972}. It leverages mathematical relationships between conditions and generation targets, allowing certain conditions to be introduced during generation through mathematical derivation without altering the training process. Thus, it significantly reduces the cost of network training.
Meanwhile, we can also directly incorporate the conditional information into the denoising network's training to enhance its ability to handle specific tasks \cite{mohsin2025conditionalpriorbasednonstationarychannel, yang2025diffusionmodelswirelesstransceivers, 11142589}. Typically, it requires additional training effort but can effectively process mathematically inexpressible conditions and other complex conditional information.

Currently, most channel estimation methods \cite{10446413, 9957135,10930691,10946972} adopt the first approach that trains without conditions. However, since in complex scenarios involving multiple antennas, multiple carriers, multiple users, and other complicated conditions, transforming the problem into an inverse problem may incur high computational costs or may not even be feasible. Thus, we aim to further explore the potential of conditionally trained models in channel estimation, where we directly train networks to adapt to these complex conditions. Although requiring more efforts in training, such method can provide greater  flexibility and does not bring too more complexity in deployment.
\subsection{Related Works}

\subsubsection{Compressed sensing methods for channel estimation}
Channel estimation based on compressed sensing (CS) methods typically assumes signal sparsity in the delay, doppler, and angular domains \cite{6674179}. Algorithms like LASSO \cite{7094443} achieve sparse signal or low-rank matrix recovery by relaxing the non-convex $l_0$-norm minimization into convex $l_1$-norm minimization. However, this approach of seeking global optimal solutions is computationally intensive. The greedy iterative algorithms such as orthogonal matching pursuit \cite{7458188} leverage pre-constructed dictionary matrices for sparse signal recovery. They offer lower computational complexity when the signal is sparse, but their accuracy depends heavily on initialization and stopping conditions. Message passing \cite{9805776} algorithms represent another class of sparse estimation methods. These algorithms recover sparse signals through the thresholding starting from noisy measurements, exemplified by the generalized approximate message passing (AMP) \cite{6033942} algorithm to achieve high-dimensional channel estimation and memory AMP \cite{9805776} for large-scale systems.

\subsubsection{DL-based channel estimation methods}
DL-based channel estimation methods have achieved substantial research results. \cite{8944280} and \cite{8752012} employ convolutional NNs (CNN) for channel estimation. 
\cite{9410430} and \cite{9526282} innovatively introduce the attention mechanism \cite{NIPS2017_3f5ee243} into the channel estimation tasks, utilizing attention to learn long-range dependencies after CNN captures the local information. \cite{10445518} introduces MLP-Mixer (CMixer) to reconstruct the full channel from partial subcarrier and antenna observations using two complex-domain multi-layer perceptron (MLP) modules that separately learn spatial and frequency characteristics. \cite{10845822} further incorporates past channel states, proposing the use of a transformer to learn temporal correlations in the channel to assist current channel estimation. Recent researches have also explored using generative models such as generative adversarial networks \cite{9252921} and variational autoencoders \cite{10960353} for channel estimation. These methods aim to learn the latent distribution of channel data and structural features of the channel to obtain the desired channel.

\subsubsection{DMs for channel estimation}
DMs methods based on pre-trained models have been applied to the channel estimation problem. The works in \cite{zilberstein2023solvinglinearinverseproblems,daras2024surveydiffusionmodelsinverse,meng2024diffusionmodelbasedposterior} have demonstrated both the applicability of DMs to general inverse problems and their derivation methodology. These works approximate the score function incorporating conditional information with a Gaussian distribution, thereby enabling the explicit derivation of posterior inference algorithms. \cite{9957135,10930691,10946972} model the MIMO channel estimation problem as a standard inverse problem, introducing the received signal, pilot sequences, and noise as conditions into the score function to derive sampling formulas for posterior inference. These methods essentially leverage the channel prior distribution learned by DMs without requiring additional modifications to the denoising network or training methods. The works in \cite{10446413,cai2025jointactivitydetectionchannel,yang2025generativediffusionreceiversachieving} jointly process signal estimation and data detection, allowing these two processes to provide mutual feedback during iteration to enhance performance. Different from the prior works, \cite{mohsin2025conditionalpriorbasednonstationarychannel, yang2025diffusionmodelswirelesstransceivers, 11142589} incorporate conditional information into the network training. Specifically, \cite{yang2025diffusionmodelswirelesstransceivers} introduces raw estimates at pilot positions as a condition during training and employs the repainting method \cite{Lugmayr_2022_CVPR} during inference, while \cite{11142589} takes the position of the users, which is difficult to incorporate into inverse problem modeling, as the conditional input during training to generate channel data corresponding to the specific positions.

\subsubsection{Conditional DMs for inpainting tasks}

The methods employed by DMs in inpainting tasks across other domains provide valuable insights for training conditional models. \cite{9887996,saharia2022paletteimagetoimagediffusionmodels,goffinet2025diffnmr2,yan2025diffnmr3advancingnmrresolution} concatenate mask-related information onto noised data to incorporate the conditions. \cite{ju2024brushnetplugandplayimageinpainting} and \cite{Zhang_2023_ICCV} introduce information from known regions at every layer of the UNet. \cite{Xie_2023_CVPR} utilizes different resolution masks for training and the outputs include the mask, which aims to preserve the original background surrounding information and understand the masks. \cite{Rombach_2022_CVPR} compresses images into latent space to reduce processing costs, then integrates multimodal conditional information through a cross-attention mechanism. \cite{Peebles_2023_ICCV} combines vision transformers with DMs, proposing the diffusion transformer (DiT) architecture. \cite{Lu_2025_ICCV} further optimizes the method of introducing conditional information based on DiT to address semantic inconsistency between completed content and original content. \cite{ICLR2025_096347b4} unifies denoising diffusion models with physics-informed NNs (PINNs) and informs the model of PDE constraints during training while ensuring inference remains unaffected.

\subsection{Contributions}
To acquire the high-fidelity channel based on the sparse and noisy raw estimates in MIMO-OFDM systems, and to enable the network to accommodate varying noise levels and the diverse pilot patterns, we propose a conditional diffusion-based architecture. Unlike most previous approaches that adjust posterior inference algorithms, we integrate conditional information directly into the network in the training process. During inference, the number of steps can be reduced to just a few while still achieving excellent performance, with no significant performance degradation observed at small number of inference steps. The main contributions of this paper are summarized as follows:
\begin{itemize}  
    \item We propose a conditional diffusion transformer(CDiT)-based channel estimation framework for MIMO-OFDM systems that effectively fuses conditional information. It can adapt to different sparse pilot patterns and maintain robustness to varying noise levels by proper noise embeddings and weighted masking.
    \item We design and implement the proposed framework, in which the conditions are properly embedded and parsed into the diffusion blocks through cross-attention modules to align the limited noisy channel observations with the intermediate denoised channel at each time step of the generation process.
    \item We evaluate the channel acquisition accuracy and robustness through extensive simulation experiments in multiple use cases. Additionally, we verify that our model can achieve high performance with only a few inference steps, and ablation studies validate the superiority and effectiveness of the network.
\end{itemize}

\vspace{-0.5cm}
\section{Preliminaries}
In this section, we briefly introduce the basic principle of the denoising diffusion probabilistic models (DDPM), for self-containedness and continuity of symbolization \cite{NEURIPS2020_4c5bcfec}. 

DMs are one of the most powerful generative models. They learn the distribution of the data in the training set, and during inference, it generates high-quality outputs by gradually denoising a random noise \cite{NEURIPS2020_4c5bcfec}. For the forward process, DMs define a forward process that adds Gaussian noise to the data $\textbf{H}_0 \sim q(\textbf{H}_0)$ in $T$ time steps according to a variance schedule $\beta_1, \dots ,\beta_T$ and finally transforms $\textbf{H}_0$ to white Gaussian noise. Each step in the forward process is given by
\begin{equation}
    q(\textbf{H}_t|\textbf{H}_{t-1}) = \mathcal{N}(\textbf{H}_t;\sqrt{1-\beta_t}\textbf{H}_{t-1},\beta_t\textbf{I}),
    \label{eq:DM1}
\end{equation}
where sample $\textbf{H}_t$ represents the latent variable at time step $t$, and $0<\beta_1<\dots<\beta_T<1$. The forward process admits sampling $\textbf{H}_t$ at an arbitrary time step $t$ in closed form
\begin{equation}
    q(\textbf{H}_t|\textbf{H}_0)=\mathcal{N}(\textbf{H}_t;\sqrt{\bar{\alpha}_t}\textbf{H}_0,(1-\bar{\alpha}_t)\mathbf{I}).\label{eq:DM2}
\end{equation}

For the reverse process, the DMs aim to generate the outputs starting at $p(\textbf{H}_{T})=\mathcal{N}(\textbf{H}_{T};\textbf{0},\textbf{I})$. However, due to the unknown distribution of $q(\textbf{H}_0)$, the conditional distribution $q(\textbf{H}_t|\textbf{H}_{t-1})$ is typically intractable. Therefore the DMs model the reverse process as a Markov chain through a variational approach with learned Gaussian transitions  
\begin{equation}
    p_{\boldsymbol{\theta}}(\textbf{H}_{t-1}|\textbf{H}_t)=\mathcal{N}(\textbf{H}_{t-1};\boldsymbol{\mu}_{\boldsymbol{\theta}}(\textbf{H}_t,t),\boldsymbol{\Sigma}_{\boldsymbol{\theta}}(\textbf{H}_t,t)),
    \label{eq:DM3}
\end{equation}
where $\boldsymbol{\mu}_{\boldsymbol{\theta}}(\textbf{H}_t,t)$ and $\boldsymbol{\Sigma}_{\boldsymbol{\theta}}(\textbf{H}_t,t)$ are parameterized by parameters $\boldsymbol{\theta}$. Training is performed by optimizing the usual variational bound on negative log likelihood
\begin{equation}
    \begin{aligned}
&\mathbb{E}\left[-\log p_{\boldsymbol{\theta}}(\textbf{H}_{0})\right]\leq L=\mathbb{E}_{q}\left[-\log\frac{p_{\boldsymbol{\theta}}(\textbf{H}_{0:T})}{q(\textbf{H}_{1:T}|\textbf{H}_{0})}\right]\\
&=\mathbb{E}_{q}\left[\underbrace{D_{\mathrm{KL}}(q(\textbf{H}_{T}|\textbf{H}_{0})\parallel p(\textbf{H}_{T}))}_{L_{T}}\underbrace{-\log p_{\boldsymbol{\theta}}(\textbf{H}_{0}|\textbf{H}_{1})}_{L_{0}}\right.\\
& \quad+\left.\sum_{t>1}\underbrace{D_{\mathrm{KL}}(q(\textbf{H}_{t-1}|\textbf{H}_{t},\textbf{H}_{0})\parallel p_{\boldsymbol{\theta}}(\textbf{H}_{t-1}|\textbf{H}_{t}))}_{L_{t-1}}\right].
    \label{eq:KL}
\end{aligned}
\end{equation}

The term $L_{t-1}$ trains the parameters in (\ref{eq:DM3}) to perform one reverse diffusion step, which is tractable when conditioned on $\textbf{H}_0$ because $q(\textbf{H}_{t-1}|\textbf{H}_{t},\textbf{H}_{0})$ is also Gaussian
\begin{equation}
\begin{aligned}
    q(\textbf{H}_{t-1}|\textbf{H}_{t},\textbf{H}_{0})&=\mathcal{N}(\textbf{H}_{t-1};\tilde{\boldsymbol{\mu}}_{t},\tilde{\beta}_{t}\mathbf{I}),\\
    \tilde{\boldsymbol{\mu}}_{t}&=\frac{1}{\sqrt{\alpha_{t}}}\left(\textbf{H}_{t}-\frac{1-\alpha_{t}}{\sqrt{1-\bar{\alpha}_{t}}}\boldsymbol{\epsilon}_t\right),\\
    \tilde{\beta}_{t}&=\frac{1-\bar{\alpha}_{t-1}}{1-\bar{\alpha}_{t}}\beta_{t},
    \label{eq:DM4}
\end{aligned}
\end{equation}
where $\boldsymbol{\epsilon}_t\sim\mathcal{N}(\boldsymbol{0},\textbf{I})$ is the Gaussian noise in the time step $t$, $\alpha_t\overset{\Delta}{\operatorname*{=}}1-\beta_t$ and $\bar{\alpha}_t\overset{\Delta}{\operatorname*{=}}\prod_{i=1}^t\alpha_i$.

Obviously, equation (\ref{eq:DM4}) reveals that $\boldsymbol{\mu}_{\boldsymbol{\theta}}(\textbf{H}_t,t)$ must predict $\frac{1}{\sqrt{\alpha_{t}}}\left(\mathbf{H}_{t}-\frac{1-\alpha_{t}}{\sqrt{1-\bar{\alpha}_{t}}}\boldsymbol{\epsilon}_t\right)$ given $\textbf{H}_t$ and thus it is recommended to choose the parameterization
\begin{equation}
    \boldsymbol{\mu}_{\boldsymbol{\theta}}(\textbf{H}_t,t)=\frac{1}{\sqrt{\alpha_t}}\left(\textbf{H}_t-\frac{\beta_t}{\sqrt{1-\bar{\alpha}_t}}\boldsymbol{\epsilon}_{\boldsymbol{\theta}}(\textbf{H}_t,t)\right),
\end{equation}
where $\boldsymbol{\epsilon}_{\boldsymbol{\theta}}(\textbf{H}_t,t)$ is the NN which takes the noisy $\textbf{H}_t$ and the time step $t$ as inputs and is intended to predict $\boldsymbol{\epsilon}_t$ from $\textbf{H}_t$. The simplified training objective can be derived as
\begin{equation}
    \mathcal{L}_{\mathrm{DM}}(\boldsymbol{\theta})=\mathbb{E}_{\mathbf{h}_0,\boldsymbol{\epsilon}_t,t}\left[\left\|\boldsymbol{\epsilon}_t-\boldsymbol{\epsilon}_{\boldsymbol{\theta}}(\sqrt{\bar{\alpha}_t}\mathbf{H}_0+\sqrt{1-\bar{\alpha}_t}\boldsymbol{\epsilon}_t,t)\right\|_2^2\right].
    \label{eq:original training}
\end{equation}

In order to generate the final outputs $\textbf{H}_0$ in the reverse process, we should sample $\textbf{H}_{t-1}\sim p_\theta(\textbf{H}_{t-1}|\textbf{H}_t)$, which is equivalent to computing 
\begin{equation}
    \textbf{H}_{t-1}=\frac{1}{\sqrt{\alpha_t}}\left(\textbf{H}_t-\frac{\beta_t}{\sqrt{1-\bar{\alpha}_t}}\boldsymbol{\epsilon}_\theta(\textbf{H}_t,t)\right)+\tilde{\beta}_{t}\boldsymbol{\epsilon}, \mathbf{\boldsymbol{\epsilon}}\sim\mathcal{N}(\mathbf{0},\mathbf{I}), 
\end{equation}
iteratively from $t=T$ to $t=1$.

\begin{figure*}[t]
\centering
  \includegraphics[width=0.95\linewidth]{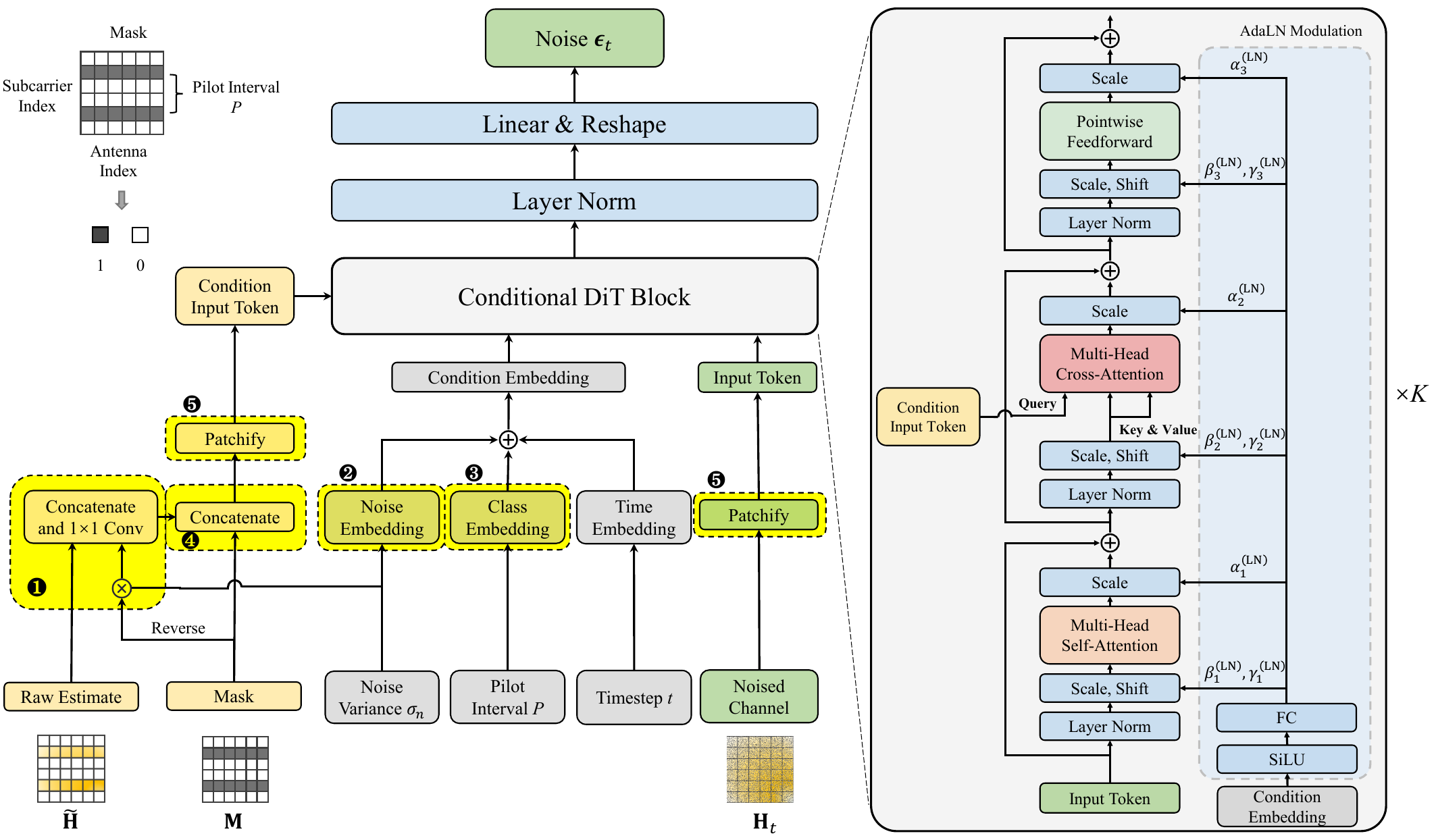}
  \caption{The proposed Conditional Diffusion Transformer (CDiT) architecture. The components highlighted in yellow boxes with numerical labels represent the operation modules, where \ding{182} denotes concatenation and $1\times1$ conv of weighted mask $\textbf{M}$, \ding{183} denotes noise embedding, \ding{184} denotes class embedding, \ding{185} denotes concatenation of mask $\textbf{M}$ and \ding{186} denotes patchify. }
  \label{fig:CDiT model}
\end{figure*}

\section{System Model}
In this paper, we consider a MIMO-OFDM system, where a base station (BS) equipped with $N_\mathrm{r} \gg 1$ antennas serves a single-antenna user equipment (UE). The considered system adopts the OFDM uplink setting with $N_\mathrm{f}$ subcarriers. We denote the channel frequency response (CFR) as $\textbf{H}\in \mathbb{C}^{N_\mathrm{f}\times N_\mathrm{r}}$. The transmitted symbol vector denoted by $\textbf{x}\in \mathbb{C}^{N_\mathrm{f}}$ is carried on the $N_\mathrm{f}$ subcarriers. Some subcarriers among all the $N_\mathrm{f}$ are selected for pilot symbols and others are for data symbols. Here we denote the set of the indices corresponding to the pilot symbols and the data symbols as $\boldsymbol{\mathcal{P}}$ and $\boldsymbol{\mathcal{D}}$, respectively. The pilot symbols $x_i, i\in \boldsymbol{\mathcal{P}}$, are all with unit power and known by the receiver. Meanwhile, the data symbols $x_i, i\in \boldsymbol{\mathcal{D}}$, are chosen from a specific constellation. Thus, the received signals denoted by $\textbf{Y}\in\mathbb{C}^{N_\mathrm{f}\times N_\mathrm{r}}$ for the specific symbol $x_k, k\in\{\boldsymbol{\mathcal{P}}, \boldsymbol{\mathcal{D}}\}$ can be expressed as
\begin{equation}
    \textbf{y}_k = x_k\textbf{h}_k + \textbf{n}_k,
    \label{eq:channel model1}
\end{equation}
where $\textbf{y}_k$ represents the $k$-th row of $\textbf{Y}$ corresponding to the $k$-th subcarrier, $\textbf{h}_k$ represents the $k$-th row of $\textbf{H}$ and $\textbf{n}_j\sim \mathcal{CN}(0, \sigma_{\mathrm{n}}^2\textbf{I})$ is a Gaussian noise. We use $\odot$ to describe the row-wise multiplication such that (\ref{eq:channel model1}) can be written as 
\begin{equation}
    \textbf{Y}=\textbf{H}\odot\textbf{x}+\textbf{N},
    \label{eq:channel model2}
\end{equation}
where \textbf{N} is the concatenation of $\textbf{n}_j$.
Based on the known pilot symbols, the initial channel response at pilot positions  $\widetilde{\textbf{h}}_i,i\in\boldsymbol{\mathcal{P}}$, can be computed as $ \widetilde{\textbf{h}}_i = {\textbf{y}_i}/{x_i}=\textbf{h}_i+\widetilde{\textbf{n}}_i,$ where $\widetilde{\textbf{n}}_i={\textbf{n}_i}/{x_i}\sim \mathcal{CN}(0, \sigma_{\mathrm{n}}^2\textbf{I})$ because $|x_i|^2=1$. 

Therefore, we aim to recover $\textbf{H}$ based on the observed noisy channel on partial subcarriers, which corresponds to the raw estimates $\widetilde{\textbf{h}}_i$ at the pilot positions and can be modeled as 
\begin{equation}
\begin{aligned}
         \mathrm{g}_0: \left [\widetilde{\textbf{h}}_i \right ]_{i\in\boldsymbol{\mathcal{P}}} \in \mathbb{C}^{\left | \boldsymbol{\mathcal{P}}\right |\times N_\mathrm{r}}  &\to \textbf{H}\in \mathbb{C}^{N_\mathrm{f}\times N_\mathrm{r}}, \\
\end{aligned}
    \label{eq:origin channel model}
\end{equation}
where $\left|\boldsymbol{\mathcal{P}}\right|$ represents the number of elements of $\boldsymbol{\mathcal{P}}$. 

The pilot positions in a single observation correspond to a specific pilot pattern, which is known by the receiver. We represent the pilot pattern using the mask $\textbf{M}\in\{0,1\}^{N_\mathrm{f}\times N_\mathrm{r}}$ where 1 indicates the positions of pilot symbols and 0 indicates the positions of data symbols. $\textbf{M}$ is a striped matrix with all ones in $i$-th rows where $i\in\boldsymbol{\mathcal{P}}$ and zeros elsewhere. The channel estimation problem in (\ref{eq:origin channel model}) can be transformed to the following mathematical model:  
\begin{equation}
\begin{aligned}
         \mathrm{g}: \widetilde{\textbf{H}} \in \mathbb{C}^{N_\mathrm{f}\times N_\mathrm{r}}  &\to \textbf{H}\in \mathbb{C}^{N_\mathrm{f}\times N_\mathrm{r}}, \\
     \widetilde{\textbf{H}} &= (\textbf{H} + \widetilde{\textbf{N}})\otimes\textbf{M},
\end{aligned}
\label{eq:channel model3}
\end{equation}
where $\otimes$ represents the element-wise multiplication and $\widehat{\textbf{N}}\sim \mathcal{CN}(0, \sigma_n^2\textbf{I})$ represents the noise. Although the noise variance at non-pilot positions is not necessarily $\sigma_{\mathrm{n}}^2$, the non-pilot components do not affect our solution after multiplying the CFR matrix by $\textbf{M}$. Thus, for simplicity, we use $\widetilde{\textbf{N}}$ as a unified representation for the noise.

Although the modeling described above only involves a single UE, it is still applicable to multi-user or multi-stream MIMO scenarios. Different users or data streams correspond to distinct pilot patterns. In multi-user systems, we typically employ non-overlap pilot patterns to obtain raw channel estimates corresponding to different users or streams independently.
Moreover, since these conditions are typically satisfied in practical systems, we simply need to treat the channel corresponding to each user or data stream as an independent sample and realize multi-user or multi-stream MIMO channel estimation through parallel computing.
\section{Proposed Methods}

In this section, we will introduce the proposed framework and its design and implementation in detail.
\subsection{Motivations on the Framework Design}
In practical systems, various pilot insertion patterns are typically adopted to accommodate dynamically changing communication scenarios. These include different pilot intervals, and even when the pilot interval is the same, the starting positions of pilots for different users may differ in multi-user scenarios as well as the noise levels. However, unlike the image inpainting tasks, the partial initial estimates we obtain are usually noisy rather than clean. Therefore, the channel estimation problem describe in (\ref{eq:channel model3}) should be modeled as an inpainting task based on the partial noisy estimates. This requires our network to not only possess denoising capabilities, but also to understand the pilot insertion patterns, and more importantly, the structural information of the wireless channels. 

Although numerous studies have demonstrated the strong capability of CNNs in image processing, \cite{10061451} points out that the channel estimation is not fully equivalent to the image inpainting task. This is because wireless channels have the strict structural properties. The relative positional information of each element plays a crucial role in reflecting this structure; even elements that are far apart can still be strongly correlated. However, CNNs are insensitive to the global positional information, making them not the most ideal choice for processing large-scale channel data. We find that the U-Net following the backbone of PixelCNN++ \cite{NEURIPS2020_4c5bcfec} which is widely used in the DMs for the imaging processing fails to achieve satisfactory results in the experiment. \cite{10445518} and \cite{10845822} have demonstrated the promising capabilities of fully connected networks and transformers in channel estimation. However, as the number of carriers and antennas increases, CDNet incurs substantial computational costs, especially CDNet due to its transformer-based implementation that has quadratic cost in the number of elements. This motivates us to investigate transformer-based solutions that can effectively learn global channel characteristics while maintaining computational efficiency.

DiT \cite{Peebles_2023_ICCV} offers effective solutions to the aforementioned challenges. It employs a patch-based tokenization approach and operates transformer on the sequences of patches. However, the original DiT model fails to meet the requirements for adapting to dynamic communication scenarios, as the network cannot generate the desired channel matrices without proper guidance. To address this, we proposed the framework of CDiT as shown in Fig. \ref{fig:CDiT model} and detailed in the following, which introduces conditional information via cross-attention in the diffusion transformer blocks with proper prompt input embeddings during model training, enabling the network to perceive both the various pilot patterns and noise in the raw estimates, and align the raw estimates with intermediate channel samples during the generation.  

\subsection{The Proposed CDiT-based Channel Estimation Framework}
Our proposed channel estimation framework is shown in Fig. \ref{fig:CDiT model}. Given that CFR is complex-valued, we separate its real and imaginary components and stack them along a new dimension. Thus both of the shape of the noised channel matrices ${\textbf{H}}_t$ at time step $t$ and the raw estimates $\widetilde{\textbf{H}}$ are ${2\times N_\mathrm{f}\times N_\mathrm{r}}$. Here, the "Noised Channel" ${\textbf{H}}_t$ we refer to corresponds to the latent variable at timestep $t$ in the standard DMs. During the training phase, it represents the channel obtained by adding noise corresponding to time step $t$ to the ground truth sample ${\textbf{H}}_0$. During the inference phase, it denotes the channel generated by denoising from Guassion noise to time step $t$. Both physically represent noise-corrupted channels, thus we uniformly use the term "Noised Channel" to describe this variable.

\subsubsection{Weighted Mask}
To enable the model to effectively comprehend diverse pilot patterns and varying noise levels during the generation of desired channels, we condition the network on the raw estimates $\widetilde{\mathbf{H}}$, the pilot mask $\mathbf{M}$, and the statistical characteristics. 
As the pilots always have normalized power, the error in the raw channel estimation shares the same variance $\sigma_{\mathrm{n}}^2$ with the channel noise.
Consequently, the noise variance $\sigma_{\mathrm{n}}^2$ serves as a critical indicator to characterize the noise level and the overall reliability of the raw estimation.

Guided by this observation, we implement a noise-weighting scheme to explicitly inform the model about the input quality across different regions. Specifically, the noise variance $\sigma_{\mathrm{n}}^2$ is multiplied element-wise with the reverse mask $\textbf{1}-\mathbf{M}$ to reflect the noise intensity relative to the pilot pattern \cite{goffinet2025diffnmr2,yan2025diffnmr3advancingnmrresolution}. 
This weighted mask then undergoes a $1 \times 1$ convolution with the raw estimates $\widetilde{\textbf{H}}$, and the resulting features are concatenated channel-wise with the original pilot mask $\textbf{M}$. 
This design ensures the network can explicitly distinguish input quality across different spatial and frequency regions. 
By internalizing the noise intensity at pilot positions, the model gains a deeper understanding of the overall reliability of the input features. This unified integration of scalar and structural conditions allows the network to adaptively interpret the "prompt" provided by sparse pilots, regardless of the noise level.

\subsubsection{Patchify}
 \begin{figure}
    \centering
    \includegraphics[width=1\linewidth]{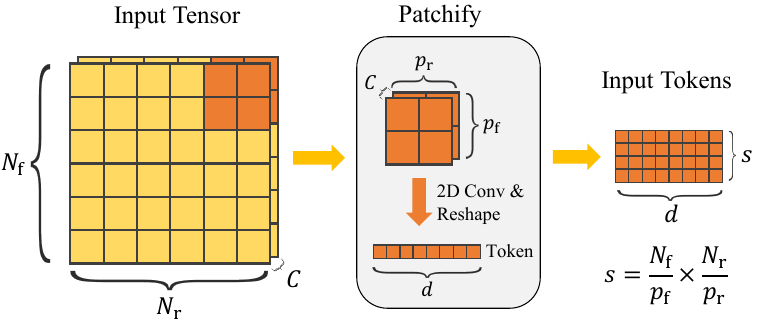}
    \caption{The illustration of patchify.}
    \label{fig:patchify}
    \vspace{-0.5cm}
\end{figure}
We implement patchify \cite{Peebles_2023_ICCV} via 2D convolution to learn intra-patch features between subcarriers and antennas. Setting both the kernel size and the stride size to $p_{\mathrm{f}}\times p_{\mathrm{r}}$, the dimension of the output channels corresponds to the embedding size of the tokens, denoted as $d$. Thus, we can transform an input tensor of dimensions ${C\times N_\mathrm{f}\times N_\mathrm{r}}$ into $s \times d$, where the sequence length $s = (N_{\mathrm{f}}/p_{\mathrm{f}})\times (N_{\mathrm{r}}/p_{\mathrm{r}})$. The schematic illustration is shown in Fig. \ref{fig:patchify}.  Additionally, we add positional encoding using sine and cosine functions into the patch sequence, which can be expressed as
 \begin{equation}
     \begin{aligned}
    PE(\text{pos}, 2j) &= \sin\left( \frac{\text{pos}}{10000^{2j/d}} \right),  \\
    PE(\text{pos}, 2j+1) &= \cos\left( \frac{\text{pos}}{10000^{2j/d}} \right),
    \label{eq:PE}
    \end{aligned}
 \end{equation}
where $\text{pos}$ denotes the position of the token within the sequence starting from index zero, and $PE(\text{pos},2j)$ is the positional encoding at position pos for the $2j$-th dimension. The variable $j$ serves as the dimension index, such that the sine function is applied to the even dimension $2j$ and the cosine function is applied to the odd dimension $2j+1$.

We perform patchify independently for $\widetilde{\textbf{H}}$ and the conditional matrices obtained from $\{\widetilde{\textbf{H}}, \textbf{M}, \sigma_{\mathrm{n}}\}$ as the dimensions of the inputs are different. What's more, we believe the CNNs extract features from these two sets of data with different focuses: the CNN processing the former focuses on extracting features inherent to the channel itself, while the latter emphasizes extracting features related to the pilot patterns and noise levels. The positional encoding used for the sequences of patches is identical, ensuring that the features from both groups remain spatially aligned.   

\subsubsection{Condition embedding}
In mainstream DMs, the time step $t$ is a pivotal component that indicates the quality (SNR) of the current diffusion variable. Inspired by this mechanism, we incorporate the statistical characteristics of the raw estimate $\widetilde{\textbf{H}}$ into the model alongside the time step $t$. 
Thus, we integrate the noise variance $\sigma_n^2$ and pilot interval $P$ as quality indicators into the model alongside the time step $t$. 

To integrate the noise variance $\sigma_n^2$, we apply the same embedding method as described in (\ref{eq:PE}), where the variable pos is substituted by $t$ and $\sigma_{\mathrm{n}}^2$, respectively, and then align their feature dimensions to $d$ via a MLP. 
To integrate the pilot interval $P$, we interpret the pilot interval as a class label and apply the classifier-free guidance \cite{ho2021classifierfree} method with a class embedding, where each specific pilot interval corresponds to a distinct class. The number of such classes is limited, as we do not consider excessively sparse pilot densities, since such extreme sparsity would make it virtually impossible to achieve channel estimation results meeting the communication requirements. It is worth mentioning that we conduct random dropout in the class label of $P$ by replacing it with a learned 'null' embedding $\emptyset$ during training. Specifically, let $N_{\mathrm{cls}}$ denote the number of possible pilot intervals. These pilot intervals are mapped to class numbers $0, 1, \dots, N_{\mathrm{cls}}-1$ in ascending order. Additionally, we introduce an extra class, mapped as $N_{\mathrm{cls}}$, to represent $\emptyset$. During training, the class number corresponding to a sample's pilot interval is replaced by $N_{\mathrm{cls}}$ with an unconditional training probability $p_{\mathrm{uncond}}$. Classifier-free guidance can be used to encourage the sampling procedure to find $\textbf{H}$ such that $\mathrm{log}p(\textbf{H}|P)$ is high. According to Bayes rule, $\nabla_{\textbf{H}} \log p(P|\textbf{H}) \propto \nabla_{\textbf{H}} \log p(\textbf{H}|P) - \nabla_{\textbf{H}} \log p(\textbf{H})$. Thus, the sampling can be performed by the linear combination of the class conditional and unconditional score estimates
 \begin{equation}
     \widehat{\boldsymbol{\epsilon}}_{\boldsymbol{\theta}}(\textbf{H}_t, P, \gamma) \propto  \boldsymbol{\epsilon}_{\boldsymbol{\theta}}(\textbf{H}_t, \emptyset) + \gamma \cdot (\boldsymbol{\epsilon}_{\boldsymbol{\theta}}(\textbf{H}_t, P) - \boldsymbol{\epsilon}_{\boldsymbol{\theta}}(\textbf{H}_t, \emptyset)),
 \end{equation}
where $\gamma$ is the guidance strength. This approach is widely-known as an effective method for improving the performance of DMs \cite{ho2021classifierfree}. Finally, all of the embeddings are added to the raw input, and the shape of the condition embedding is $1 \times d$.
\subsubsection{Conditional diffusion transformer block}
After patchify, the input tokens and conditional information pass through a sequence of transformer blocks. Following the DiT model design, each transformer block includes a self-attention module and a point-wise feedforward module. Additionally, we continue to employ adaptive layer normalization (adaLN) which takes condition embedding as input to process the scalar conditional information \cite{Peebles_2023_ICCV}. This replaces the standard layer norm layers in transformers.

\begin{algorithm}[t!] 
  \caption{Training Algorithm}
  \label{alg:training}
  \begin{algorithmic}[1]
    \REQUIRE $p_{\mathrm{uncond}}$, SNR range $[r_{\mathrm{min}},r_{\mathrm{max}}]$
    \REPEAT
      \STATE Sample $\textbf{H}_0 \sim q(\textbf{H}_0)$ 
      \STATE Sample power $P_{\mathrm{signal}}=\|\textbf{H}_0\|_\mathrm{F}^2/(\boldsymbol{N}_\mathrm{r}\boldsymbol{N}_\mathrm{f})$
      \STATE $\textbf{H}_0 = \textbf{H}_0/\sqrt{P_{\mathrm{signal}}}$
      \STATE Sample $t \sim \text{Uniform}(\{1, \dots, T\})$         
      \STATE Sample $\boldsymbol{\epsilon} \sim \mathcal{N}(0, \mathbf{I})$  
      \STATE Sample $\textbf{M}$ from all the possible pilot patterns with pilot interval $P$
      \STATE $P\leftarrow \emptyset$ with probability $p_{\mathrm{uncond}}$
      \STATE Sample $r$ from $[r_{\mathrm{min}}, r_{\mathrm{max}}]$, $\sigma_{\mathrm{n}}^2 = P_{\mathrm{signal}}/r$, sample $\textbf{N}\sim \mathcal{CN}(0, \sigma_{\mathrm{n}}^2\textbf{I})$
      \STATE $\widetilde{\textbf{H}}=(\textbf{H}_0+\textbf{N})\otimes\textbf{M}$
      \STATE $\boldsymbol{c}\leftarrow\{{\widetilde{\textbf{H}}}, \textbf{M}, \sigma_{\mathrm{n}}, P, p_{\mathrm{uncond}}\}$
      \STATE Take gradient descent step on $\nabla_{\boldsymbol{\theta}} \left\|\boldsymbol{\epsilon} - \widehat{\boldsymbol{\epsilon}}_{\boldsymbol{\theta}}(\sqrt{\bar{\alpha}_t}\textbf{H}_0 + \sqrt{1 - \bar{\alpha}_t}\boldsymbol{\epsilon}, t, \boldsymbol{c})\right\|^2$  
    \UNTIL converged                                                
  \end{algorithmic}
\end{algorithm}

\begin{algorithm}[t!]
  \caption{Sampling Algorithm}
  \label{alg:sampling}
  \begin{algorithmic}[1]
  \REQUIRE $\gamma$, SNR $r$, $\widetilde{\textbf{H}}$, $\textbf{M}$, $P$, $\boldsymbol{\tau}$, $\eta$
    \STATE Sample $\textbf{H}_T \sim \mathcal{N}(0, \mathbf{I})$  
    \STATE $\textbf{H}_{\boldsymbol{\tau}_{S}} = \textbf{H}_T$
    \STATE Power of raw estimates $P_{\mathrm{signal}} = \|\widetilde{\textbf{H}}\|_\mathrm{F}^2/\|\textbf{M}\|_1$
    \STATE $\widetilde{\textbf{H}} = \widetilde{\textbf{H}} \times\frac{1}{\sqrt{P_{\mathrm{signal}}}}\times \sqrt{\frac{r+1}{r}}$
    \FOR{$s = S, \dots, 1$}                                           
      \STATE Sample $\boldsymbol{\epsilon} \sim \mathcal{N}(0, \mathbf{I})$ if $\boldsymbol{\tau}_s > 1$, else $\boldsymbol{\epsilon} = 0$ 
      \STATE $\sigma_{\mathrm{n}}^2 = \frac{1}{r+1} \times\frac{1}{\|\textbf{M}\|_1}\|\widetilde{\textbf{H}}\|_\mathrm{F}^2 =1/r$
      \STATE $\boldsymbol{c}\leftarrow\{{\widetilde{\textbf{H}}}, \textbf{M}, \sigma_{\mathrm{n}}, P, p_{\mathrm{uncond}}\}$
      \STATE Calculate $\textbf{H}_{\boldsymbol{\tau}_{s-1}}$ according to (\ref{eq:DDIM})
    \ENDFOR                  
    \STATE $\widehat{\textbf{H}}_0 = \textbf{H}_0 \times \sqrt{P_{\mathrm{signal}}}\times\sqrt{\frac{r}{r+1}}$
    \STATE Return $\widehat{\textbf{H}}_0$                                        
  \end{algorithmic}
\end{algorithm}

Instead of directly concatenating the conditional information at the input \cite{yan2025diffnmr3advancingnmrresolution, 9887996}, we apply the cross-attention module to incorporate the conditional input tokens. Here, the conditional input token serves as the query, while the output from the previous self-attention acts as the key and the value, respectively. This design choice is driven by the need for deep, prompt-aware fusion, where the conditions can effectively modulate the feature representation. The cross-attention mechanism explicitly models the correlation between the conditions and the latent features of the noised channel, allowing the network to adaptively attend to the most relevant conditional information based on the current context. Furthermore, we apply the adaLN for the cross-attention module. Thus we have three groups of dimension-wise scale and shift parameters $\{\alpha_{i}^{\mathrm{(LN)}}, \beta_{i}^{\mathrm{(LN)}}, \gamma_{i}^{\mathrm{(LN)}} \}$, where $i =1, 2, 3$ and the shape of the parameters is $1\times d$. The fully connected layer in adaLN modulation transforms the shape of the condition embedding from $1\times d$ to $9\times d$. We denote the number of CDiT block is $K$. 

\section{Training and Improving Inference Speed \label{section:improve speed}}

\subsection{Training Method}
The amplitude of the CFR data exhibits significant variations. However, DMs typically require inputs normalized within the range of $[-1, 1]$ for better performance. Thus we normalize the data by dividing each sample $\mathbf{H}_0$ by a power factor which can be derived as $\|\textbf{H}_0\|_\mathrm{F}^2/(\boldsymbol{N}_\mathrm{r}\boldsymbol{N}_\mathrm{f})$, where $\|\cdot\|_\mathrm{F}$ represents the Frobenius norm. This ensures that the input amplitudes remain approximately on the order of 1, resulting in a more uniform sample distribution and effectively reducing the learning difficulty for the network. During inference, it is necessary to estimate the power of the CFR to rescale the network's output and obtain the final results. Since only noisy raw estimates at the pilot positions are available, we calculate the power of CFR, which can be derived as $P_{\mathrm{signal}} = \|\widetilde{\textbf{H}}\|_\mathrm{F}^2/\|\textbf{M}\|_1$, where $\|\cdot\|_\mathrm{1}$ represents the element-wise $l_1$-norm. This value is then used to derive the power of the sample with the SNR. We assume that the SNR is knwon for each sample. This approach is consistent with the power calculation method used during the training, with the distinction that the inference relies on partial observations to estimate the power.

After the pre-processing described above, we can obtain a certain quantity of accurate channel $\textbf{H}$. During training, we perform mixed training with noise of varying variances and different pilot patterns. Specifically, before the sample is fed into the network, we randomly generate noise and pilot patterns. We use $\widetilde{\textbf{H}}$ as the raw estimates. After incorporating conditional information $\{{\widetilde{\textbf{H}}}, \textbf{M}, \sigma_{\mathrm{n}}, P, p_{\mathrm{uncond}}\}$ denoted as $\boldsymbol{c}$ into the training, different from (\ref{eq:original training}), the training objective of our CDiT can be expressed as
\begin{equation}
     \mathcal{L}_{\mathrm{DM}}(\boldsymbol{\theta}) = \mathbb{E}_{\mathbf{h}_0,\boldsymbol{\epsilon}_t,t}\left[\left\|\boldsymbol{\epsilon}_t-\widehat{\boldsymbol{\epsilon}}_{\boldsymbol{\theta}}(\sqrt{\bar{\alpha}_t}\mathbf{H}_0+\sqrt{1-\bar{\alpha}_t}\boldsymbol{\epsilon}_t,t, \boldsymbol{c} )\right\|_2^2\right],    
\end{equation}
The training algorithm is shown in Algorithm \ref{alg:training}.

\subsection{Sampling Method}
 \begin{figure*}[t!]
    \centering
    \includegraphics[width=0.8\linewidth]{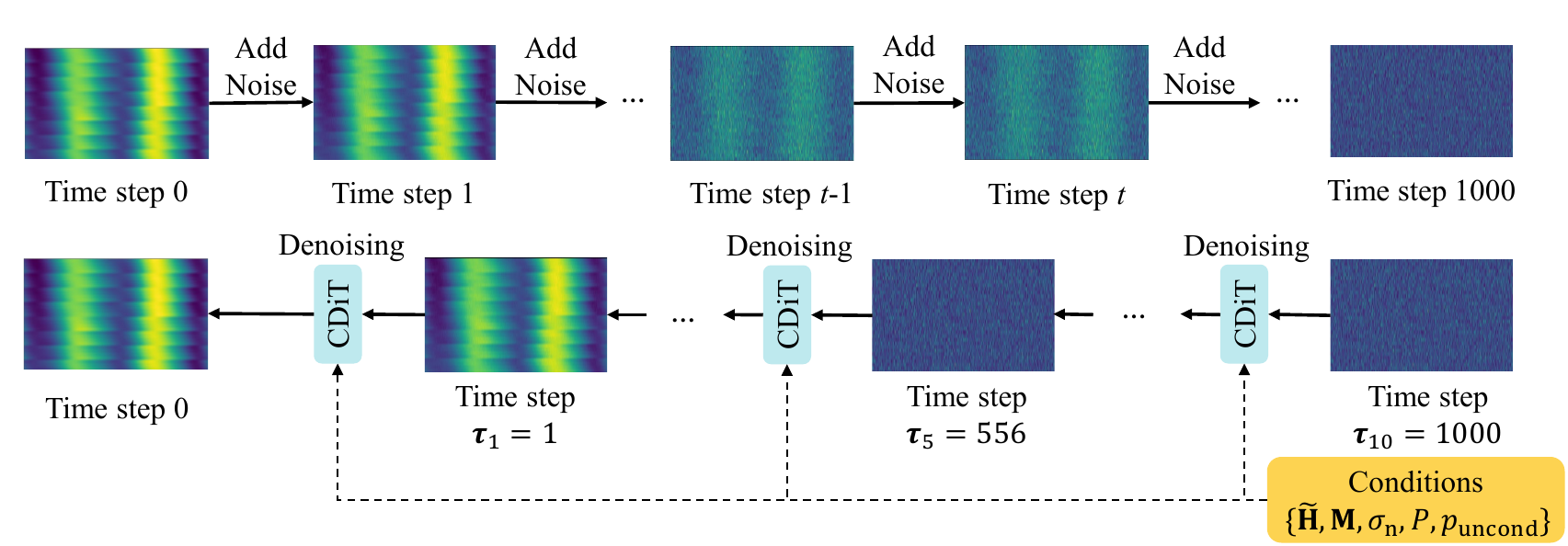}
    \caption{The illustration of the forward process and reverse sampling process of the proposed framework. The number of time steps in training is 1000. We select a subset of time steps $\boldsymbol{\tau}$ with length $S$ for inference. The illustration takes $S=10$ as an example.}
    \label{fig:training_inference_steps}
    \vspace{-0.3cm}
\end{figure*}

 All of our models are trained for 1000 diffusion steps during training. However, if the same number of inference steps are used during inference, it would result in unacceptably long inference times. Therefore, for diffusion models, it is crucial to reduce the number of inference steps while avoiding significant performance degradation.
 Thus, following the sampling method in \cite{pmlr-v139nichol21a}, we select a set of time steps $\boldsymbol{\tau}$ which is an increasing sub-sequence of $[1,\dots,T]$ with length $S$. This means that the model completes the generation process based on the subset $\{ \textbf{H}_{\boldsymbol{\tau}_1}, \dots, \textbf{H}_{\boldsymbol{\tau}_S} \}$, and the length of the sampling trajectory is much smaller than $T$. We employ linear sampling \cite{pmlr-v139nichol21a,Lin_2024_WACV} to select the inference time steps. Specifically, this method includes both the first and the last time step, namely $\boldsymbol{\tau}_1=1$ and $\boldsymbol{\tau}_S=T$, and then uniformly selects intermediate time steps via linear interpolation. The general sampling equation is given by \cite{song2020denoising}
\begin{equation}
\begin{aligned}
    &\textbf{H}_{\boldsymbol{\tau}_{s-1}}(\eta) = \sqrt{\bar{\alpha}_{\boldsymbol{\tau}_{s-1}}} \left( \frac{\textbf{H}_{\boldsymbol{\tau}_s} - \sqrt{1 - \bar{\alpha}_{\boldsymbol{\tau}_s}} \boldsymbol{\epsilon}_\theta(\textbf{H}_{\boldsymbol{\tau}_s}, \boldsymbol{\tau}_s,\boldsymbol{c})}{\sqrt{\bar{\alpha}_{\boldsymbol{\tau}_s}}} \right) \\
    & + \sqrt{1 - \bar{\alpha}_{\boldsymbol{\tau}_{s-1}} - \sigma_{\boldsymbol{\tau}_s}(\eta)^2} \cdot \boldsymbol{\epsilon}_\theta(\textbf{H}_{\boldsymbol{\tau}_s}, \boldsymbol{\tau}_s, \boldsymbol{c}) + \sigma_{\boldsymbol{\tau}_s}(\eta) {\boldsymbol{\epsilon}},
    \label{eq:DDIM}
\end{aligned}
\end{equation}
where
\begin{equation}
\sigma_{\boldsymbol{\tau}_s}(\eta) = \eta \sqrt{ \frac{1 - \bar{\alpha}_{\boldsymbol{\tau}_{s-1}}}{1 - \bar{\alpha}_{\boldsymbol{\tau}_s}}} \sqrt{1 - \frac{\bar{\alpha}_{\boldsymbol{\tau}_s}}{\bar{\alpha}_{\boldsymbol{\tau}_{s-1}}}},
\end{equation}
where $s=1,...,S$. For the boundary conditions, we denote $\boldsymbol{\tau}_0=0$ and $\bar{\alpha}_{\boldsymbol{\tau}_0}=\bar{\alpha}_{0}=1$, which is also implemented in diffusers \cite{von-platen-etal-2022-diffusers}. This means that in the last time step, we should directly output the predicted sample rather than add the random noise. The diagram of the forward process and reverse sampling process is shown in Fig. \ref{fig:training_inference_steps}. 

When $\eta=1$, the generative process becomes a DDPM, whereas when $\eta=0$, it becomes a denoising diffusion implicit models (DDIM) \cite{song2020denoising}. For DDIM, the samples are generated from latent variables through a fixed procedure because the coefficient before the random noise ${\boldsymbol{\epsilon}}$ becomes zero. 
However, in our scenario, the channel is perturbed by random noise, and the generation process is guided by the noisy raw estimates along with other conditional information. In this context, we believe that keeping the random noise term rather than modeling the sample generation as a deterministic process during inference would be more reasonable. The sampling algorithm is summarized in Algorithm \ref{alg:sampling}.

\section{Numerical Results}
In this section, we evaluate the performance of our proposed CDiT. We will first introduce the experiment settings. Then we will compare our CDiT with other methods including the accuracy of the channel estimations, generalization and robustness to the noise, different pilot patterns and the channel types. 

\subsection{Experiment Details\label{exp:experiment detials}}
 \subsubsection{Data generation}
\begin{figure}[t]
\centering
  \includegraphics[width=0.8\linewidth]{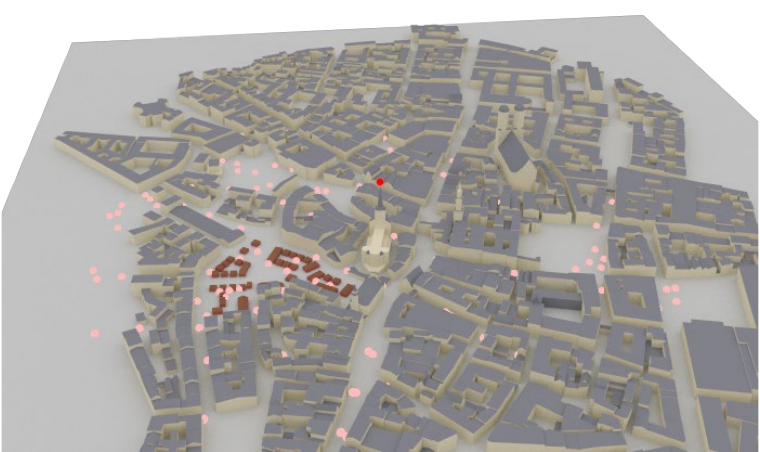}
  \caption{The scene graph of the experiments in sionna. The red dot represents the position of the BS and the pink dots represent that of the UEs.}
  \label{fig:scene}
  \vspace{-0.3cm}
\end{figure}

 For our experiments, we use sionna \cite{sionna} as the channel simulator to generate training, testing, and validation data in the scene containing the area around the Frauenkirche in Munich, Germany based on the ray-tracing, which is a basic demo scene in sionna. Fig. \ref{fig:scene} details the scene graph of the experiment. In the scenario, the red dot at the top of the tower represents the position of the BS, while the pink dots distributed throughout the urban area indicate possible positions of UEs. The BS is equipped with $4\times4$ planar antenna array with cross-polarization. Each cross-polarized antenna provides two independent channels including vertical and horizontal polarizations, hence 16 physical antennas yield 32 logical channels in the antenna dimension. The UEs are randomly distributed in the scene at straight-line distance between 100 m and 400 m from the BS, each assigned a randomly generated speed ranging from 0 to 10 m/s. At the specific UE position, we capture the uplink CFR between the UE and the BS. The dimension of the CFR data for a single UE is $14 \times 1024 \times 32$, where 14 denotes the number of OFDM symbols, each subject to a different doppler shift. We randomly select one OFDM symbol from this set to serve as the CFR sample for the corresponding UE. The channel between the UE and BS comprises multiple propagation paths, including line-of-sight, reflected, and scattered components. We set the maximum number of ray interactions as 5 and keep the 48 strongest paths according to the path gains. Due to obstructions, UEs at certain positions may have no propagation paths to the BS and thus we filter out UE positions without propagation paths. The center frequency is 3.5 GHz and the number of subcarriers is 1024. The relevant parameters of the datasets are detailed in Table \ref{tab:sionna_parameter}.

\begin{table}[!t]
  \centering
  \footnotesize
  \caption{ PARAMETER SETTINGS FOR SIONNA DATASETS}
  \begin{tabularx}{\linewidth}{
    >{\raggedright\arraybackslash\hsize=0.6\hsize}X  
    >{\raggedright\arraybackslash\hsize=0.4\hsize}X  
  }
  \toprule  
    \textbf{Parameters} & \textbf{Value} \\
    \midrule  
    Scene & Munich \\
    Center frequency & 3.5 GHz \\
    Subcarrier spacing & 15 kHz \\
    Antenna array form & UPA \\
    Polarization of antennas & Cross-polarization \\
    Number of antennas ($N_{\mathrm{r}}$) & 16 \\
    Number of subcarriers ($N_\mathrm{f}$) & 1024 \\
    Maximum number of paths & 48 \\
    Maximum number of ray interactions & 5 \\
    Position of BS & [-15, -165, 95] m \\
    Distance between UEs and the BS & 100 m $\sim$ 400 m \\
    Speed of UEs & 0 $\sim$ 10 m/s \\
    Number of training data & 8000 \\
    Number of validation data & 2000 \\
    Number of testing data &  2000 \\
    \bottomrule  
  \end{tabularx}
  \label{tab:sionna_parameter}
\end{table}

\begin{table}[!t]
  \centering
  \footnotesize
  \caption{PARAMETER SETTINGS FOR TRAINING} 
  \begin{tabularx}{\linewidth}{
    >{\raggedright\arraybackslash\hsize=0.35\hsize}X  
    >{\raggedright\arraybackslash\hsize=0.65\hsize}X  
  }
    \toprule
    Parameters & Value \\
    \midrule
    Batch size & 64 \\
    Optimizer & AdamW \\
    Base learning rate & 0.0003 \\
    LR schedule & Cosine annealing with linear warmup \\
    Warmup steps & 4500 \\
    Training epoch & 900 \\
    Time steps $T$ for training & 1000 \\
    Pilot interval $P$ & $\{2, 4, 8, 16, 32\}$ \\
    Patch size ($p_\mathrm{f},p_\mathrm{r}$) & (64, 2)\\
    Signal-to-noise $r$ & 5$\sim$35 dB \\
    Structure for CDiT &  $K=9$, $d=768$, $p_{\mathrm{uncond}}=0.1$, $\gamma=1$, $p_{\mathrm{f}}=64$, $p_{\mathrm{r}}=2$  \\
    Number of parameters & 136.52 Million \\
    Parameters for diffusion & $\beta_1=0.0001$, $\beta_{T}=0.02$, \\ 
                             & variance schedule = 'linear' \cite{Lin_2024_WACV}, \\
                             & inference time step spacing = 'linspace' \cite{Lin_2024_WACV} \\
    \bottomrule
  \end{tabularx}
  \label{tab:training_params}
  \vspace{-0.3cm}
\end{table}

\subsubsection{Performance indexes}
We use normalized MSE (NMSE) and cosine correlation $\rho$ \cite{8322184} between acquired channel and true channel as the performance indexes, which are defined as follows:
\begin{equation}
\text{NMSE} = \mathbb{E} \left\{ \frac{\|\textbf{H}_0 - \widehat{\textbf{H}}_0\|_{\mathrm{F}}^2}{\|\textbf{H}_0\|_{\mathrm{F}}^2} \right\}, 
\label{eq:nmse}
\end{equation}
and 
\begin{equation}
\rho = \mathbb{E} \left\{ \frac{1}{N_\mathrm{f}} \sum_{i=1}^{N_\mathrm{f}} \frac{\left| \widehat{\mathbf{h}}_i^\mathrm{H} \mathbf{h}_i \right|}{\| \widehat{\mathbf{h}}_i \|_2 \| \mathbf{h}_i \|_2} \right\},
\label{eq:rho}
\end{equation}
where $\mathbf{h}_i$ and $\widehat{\mathbf{h}}_i$ are the original and estimated CFR of the $i$-the subcarrier, i.e. the $i$-the rows of CFR matrices $\textbf{H}_0 $ and $ \widehat{\textbf{H}}_0$, respectively.
\subsubsection{Benchmarks and training settings}
During the training process, the SNR of the raw estimates is randomly distributed within the range of 5 to 35 dB. Additionally, the pilot symbols are equal spacing within the carrier domain. The pilot interval $P$ is randomly selected from the set $\{2, 4, 8, 16, 32\}$ and the starting position $P_{\mathrm{start}}$ is randomly chosen within the range $[0, P-1]$. Therefore, the pilot indices in the subcarrier domain are $P_{\mathrm{start}} +  [0, 1\times P,\dots, (\lfloor\frac{N_{\mathrm{f}}}{P}\rfloor-1)\times P ]$. We employ linear schedule for the variance schedule described in \cite{Lin_2024_WACV} and the method for inference time step selection is described in \ref{section:improve speed}. The relevant parameters of the training and CDiT are detailed in Table \ref{tab:training_params}. 

The following baseline methods are compared:
\begin{itemize}
    \item \textbf{Linear Interpolation}: The algorithm estimates the channels based on the raw estimates through the linear interpolation.
    \item \textbf{LMMSE}: The algorithm computes for each element of an OFDM resource grid a channel estimate and error variance through linear estimator using the sample covariance computed via all training samples. The interpolation is carried out accross frequency and antennas in order \cite{sionna}. The covariance matrices of the frequency and antennas can be calculated as $\textbf{R}_{\mathrm{f}} = \frac{1}{BN_{\mathrm{r}}} \sum_{i=1}^B {\textbf{H}_0} {\textbf{H}_0}^\mathrm{H}$ and $\textbf{R}_{\mathrm{r}} = \frac{1}{BN_{\mathrm{f}}} \sum_{i=1}^B {\textbf{H}_0}^\mathrm{H} {\textbf{H}_0}$, where $B$ is the number of samples. Computing the estimated channels across frequency as $\widehat{\mathbf{H}}_{0,\mathrm{f}} = \textbf{R}_{\mathrm{f}} \textbf{A}^\mathrm{H} (\textbf{A} \textbf{R}_{\mathrm{f}} \textbf{A}^\mathrm{H} + \sigma_\mathrm{n}^2 \textbf{A})^{-1} \widetilde{\textbf{H}}$, where $\textbf{A}=\mathrm{diag}(\textbf{m})$ and $\textbf{m}$ is a column of $\textbf{M}$. Then calculating across antennas as $\widehat{\mathbf{H}}_{0} = \textbf{R}_{\mathrm{r}}( \textbf{R}_{\mathrm{r}}+ \boldsymbol{\Sigma})^{-1} \widetilde{\textbf{H}}_{0,\mathrm{f}}$, where $\boldsymbol{\Sigma}$ is the error variance obtained from the frequency interpolation. 
    \item \textbf{CMixer}: We set the number of layers in CMixer to 8. When the pilot interval is 16, the number of parameters of CMixer is 138.90 Million. Different pilot intervals cause the model size to fluctuate by approximately 2 Million, thus ensuring that the model size remains comparable to that of CDiT. The training details and other parameters follow \cite{10445518}. 
\end{itemize}

\subsection{Performance Evaluation}
\subsubsection{Accuracy of channel estimation}

\begin{figure}[t!]
    \centering
    \includegraphics[width=0.8\linewidth]{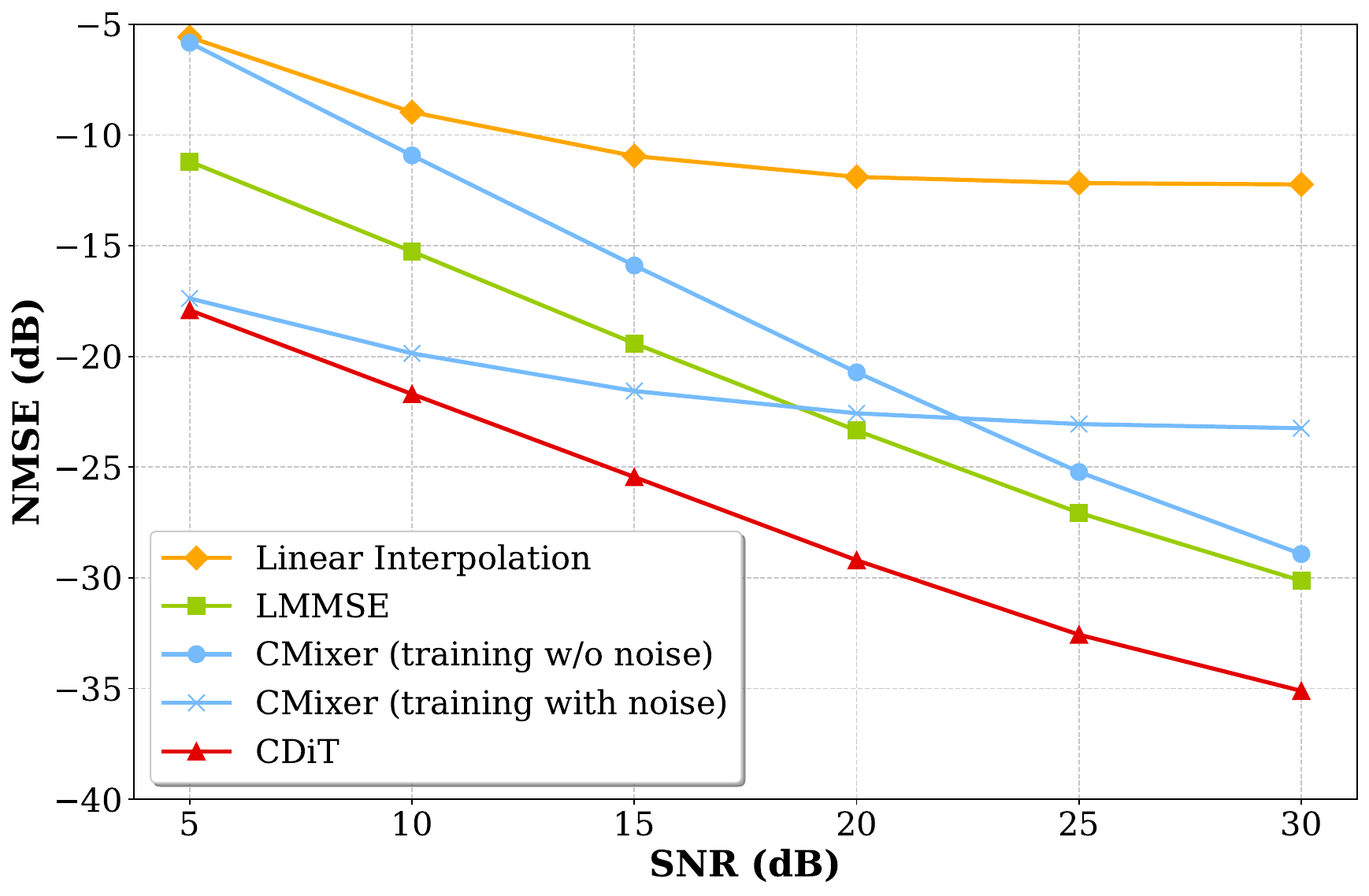}
    \caption{NMSE performance versus different SNR with $P=16$. The number of inference steps $S$ of CDiT is 10.}
    \label{fig:exp:benchmark}
\end{figure}

\begin{figure}[t!]
    \centering
    \includegraphics[width=0.8\linewidth]{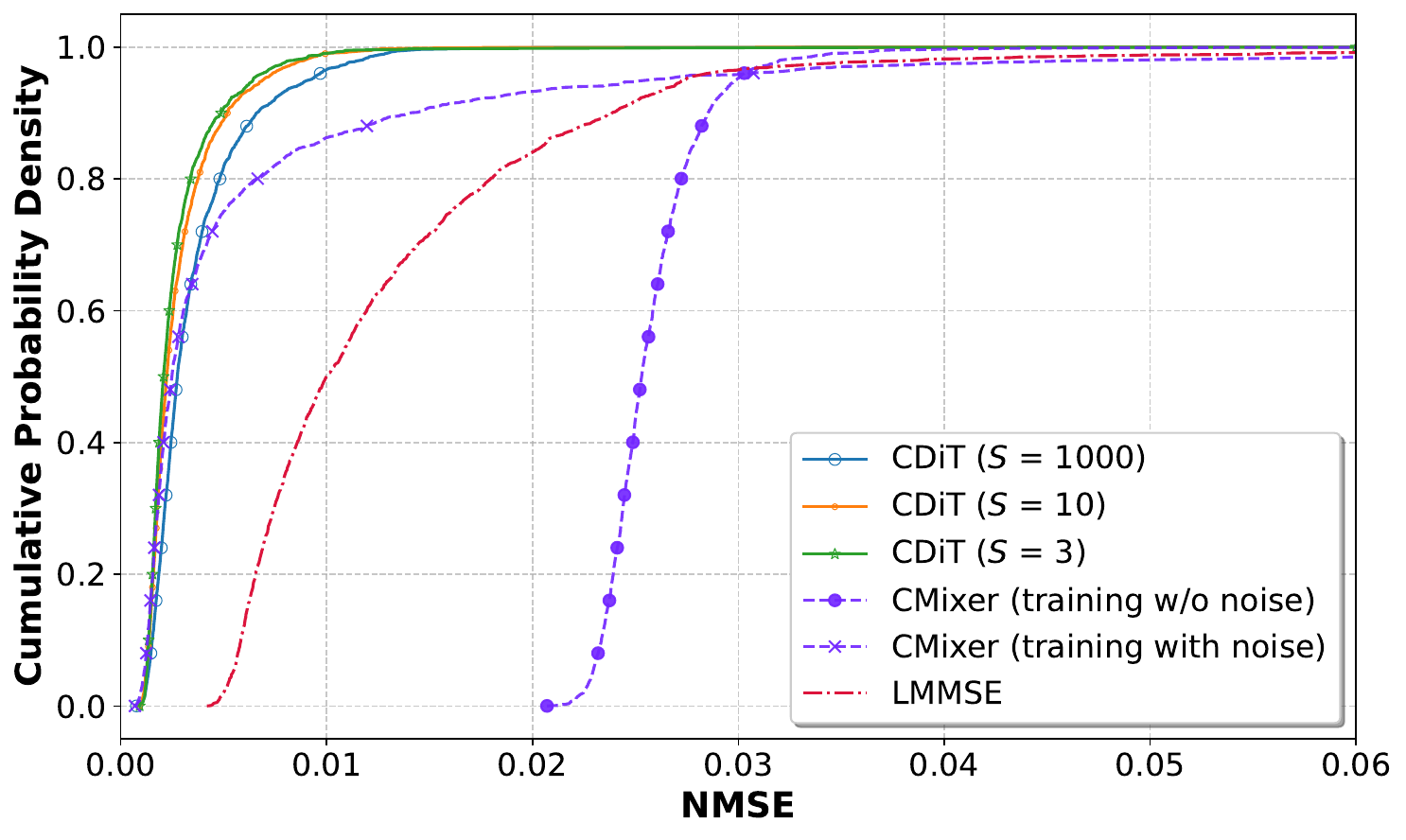}
    \caption{Cumulative probability distribution of errors (NMSE) between the estimated channel and the ground truth with $\mathrm{SNR}=15$ dB and $P=16$. }
    \label{fig:exp:cdf}
\end{figure}

\begin{figure}[t!]
    \centering
    \includegraphics[width=0.8\linewidth]{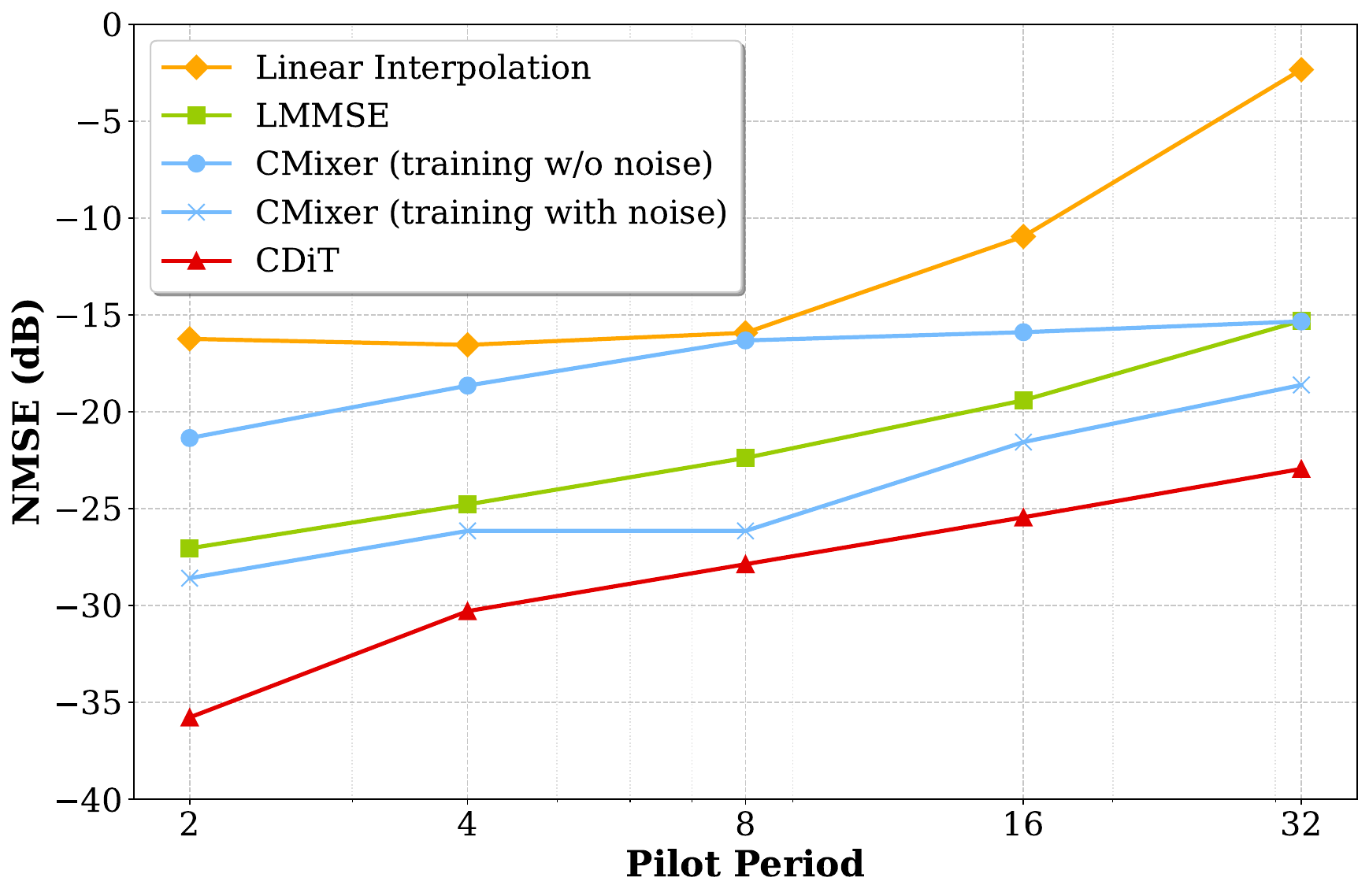}
    \caption{NMSE performance under different pilot interval with $\mathrm{SNR}=15$ dB. We set $S=10$ for CDiT.}
    \label{fig:exp:benchmark_differet_interval}
    \vspace{-0.3cm}
\end{figure}

\begin{figure}[t!]
    \centering
    \includegraphics[width=1\linewidth]{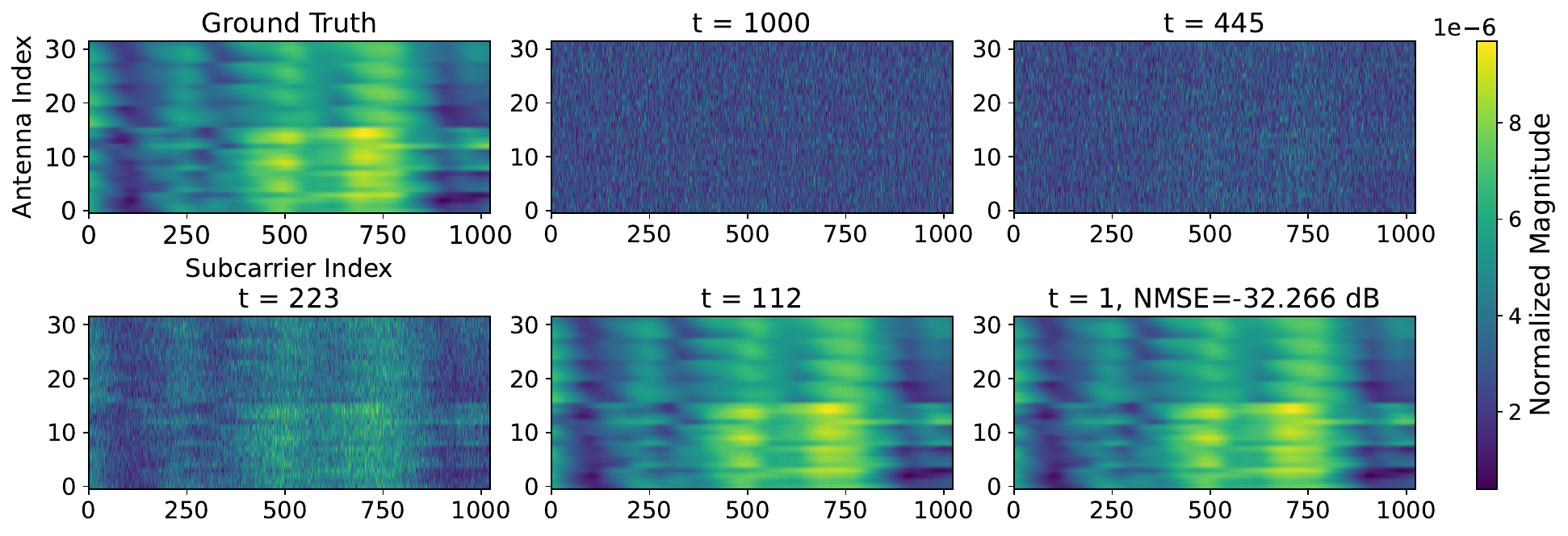}
    \caption{Channel estimation results over different time with $P=16$, $\mathrm{SNR} = 30$ dB and $S=10$.}
    \label{fig:exp:single sample show}
    \vspace{-0.3cm}
\end{figure}

We first evaluate the accuracy of the channel estimation generated by our CDiT under different conditions with $S=10$. As shown in Fig. \ref{fig:exp:benchmark}, when pilot interval is 16, our CDiT achieves strong performance under both random pilot positions and varying SNR. Since CMixer lacks the capability to adapt to random pilot configurations, only one pilot patterns are used during both training and testing for CMixer. 
When CMixer is trained in a 'w/o noise' mode, using noise-free pilot-based estimates $\textbf{H}\otimes\textbf{M}$ as the network input, its performance degrades significantly as the SNR decreases. However, when it is trained in a 'with noise' mode, using the raw estimates $\widetilde{\textbf{H}}$ as input, its performance noticeably decreases at high SNR levels. This means that CMixer does not adapt well to varying noise levels.

Fig. \ref{fig:exp:cdf} shows the cumulative probability distribution of errors between the CDiT and the benchmarks when the $\mathrm{SNR}=15$ dB and the pilot interval is 16. The cumulative probability distribution of the CDiT remains around 0.01 across different $S$. As $S$ decreases, the NMSE distribution becomes more concentrated. Fig. \ref{fig:exp:benchmark_differet_interval} shows the performance under different pilot intervals when $\mathrm{SNR}=15$ dB and $S=10$ for CDiT. The CDiT consistently has the best performance across various pilot intervals. 
\subsubsection{Performance versus inference steps $S$}

Reducing the number of inference steps is one of the most important issues in applying DMs to channel estimation. This section will show performance of the CDiT model under different inference steps. In the experiment, we set different $S$, namely $[2, 3, 4, 5, 10, 100, 200, 1000]$. Since we use a 'linspace' for time step spacing, the time steps should include both 1 and $T$, meaning the minimum steps is at least 2. One of the generation result is shown in the Fig. \ref{fig:exp:single sample show}, where we set $S$ as 10.

\begin{figure*}[t!]
  \centering
  \subcaptionbox{$P=4$}[0.32\linewidth]{%
    \centering
    \includegraphics[width=\linewidth]{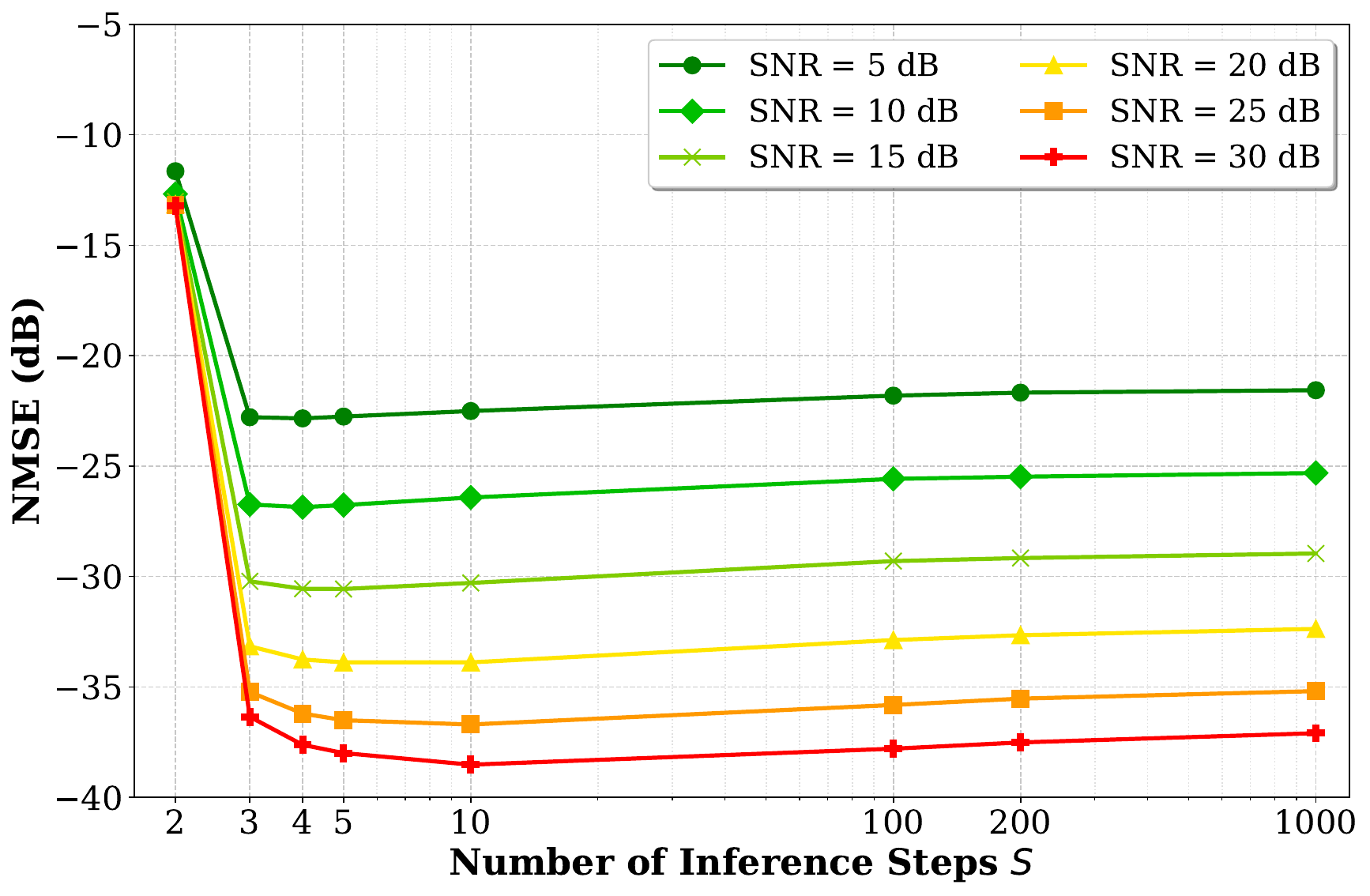}
  }%
  \hfill 
  \subcaptionbox{$P=16$}[0.32\linewidth]{%
    \centering
    \includegraphics[width=\linewidth]{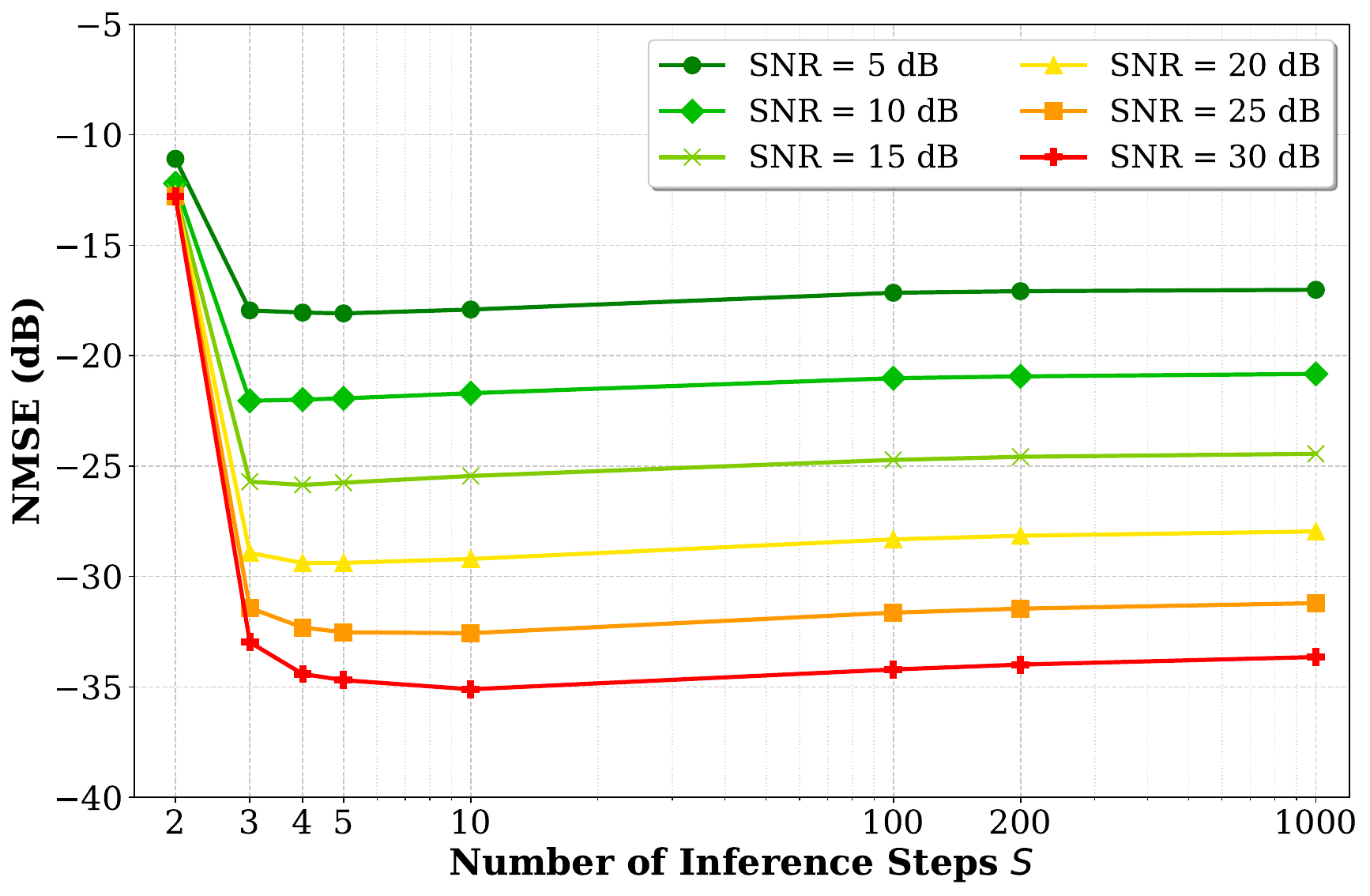}
  }%
  \hfill 
  \subcaptionbox{$P=32$}[0.32\linewidth]{%
    \centering
    \includegraphics[width=\linewidth]{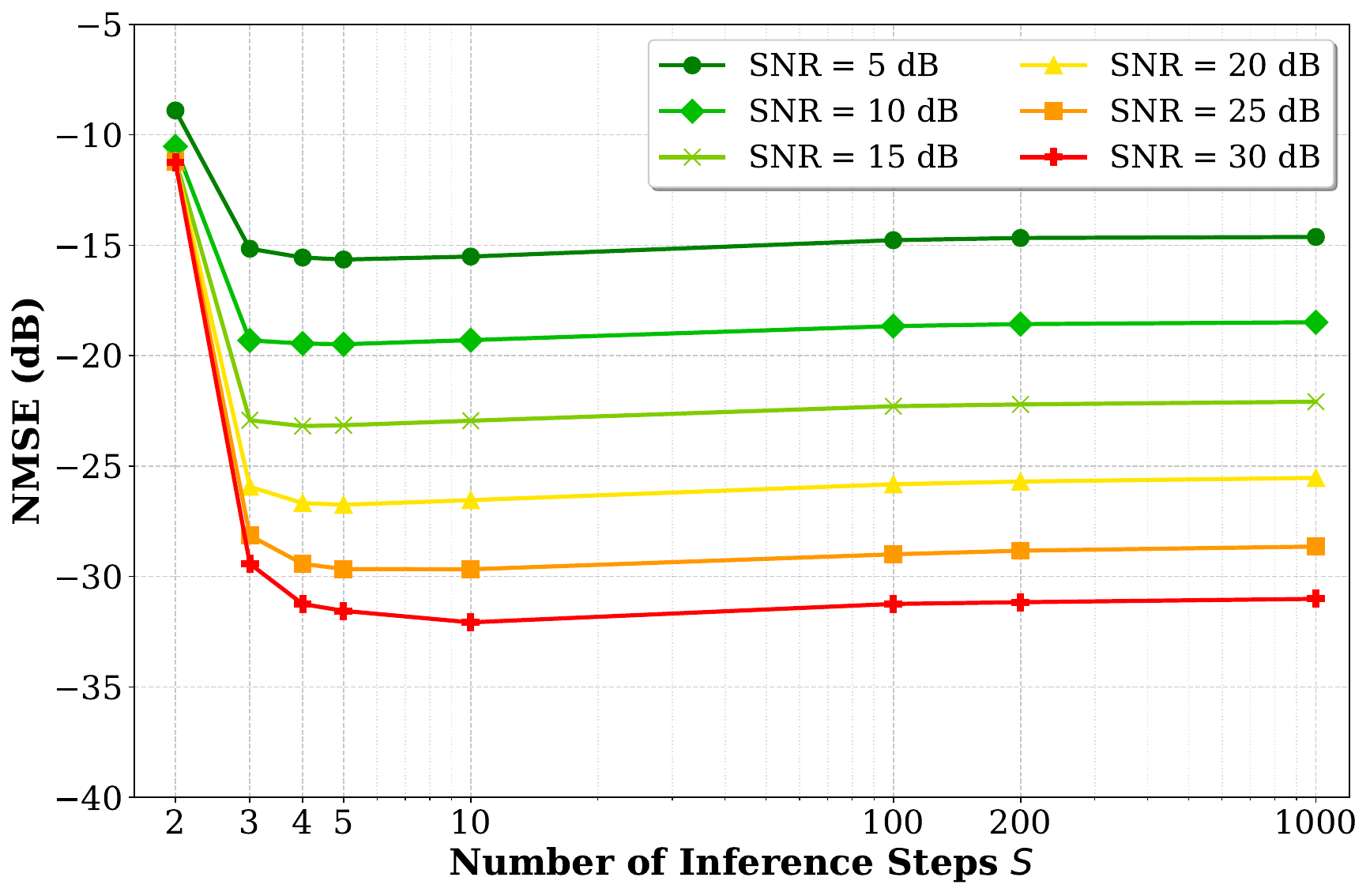}
  }%
  
  \caption{NMSE versus the number of inference steps $S$ of DDPM with 'linspace' time step spacing under different noise levels.}
  \label{fig:exp:nmse_vs_steps_ddpm}
\end{figure*}

\begin{figure*}[t!]
  \centering
  \subcaptionbox{$P=4$\label{fig:exp:nmse_vs_steps_4_ddim}}[0.32\linewidth]{%
    \centering
    \includegraphics[width=\linewidth]{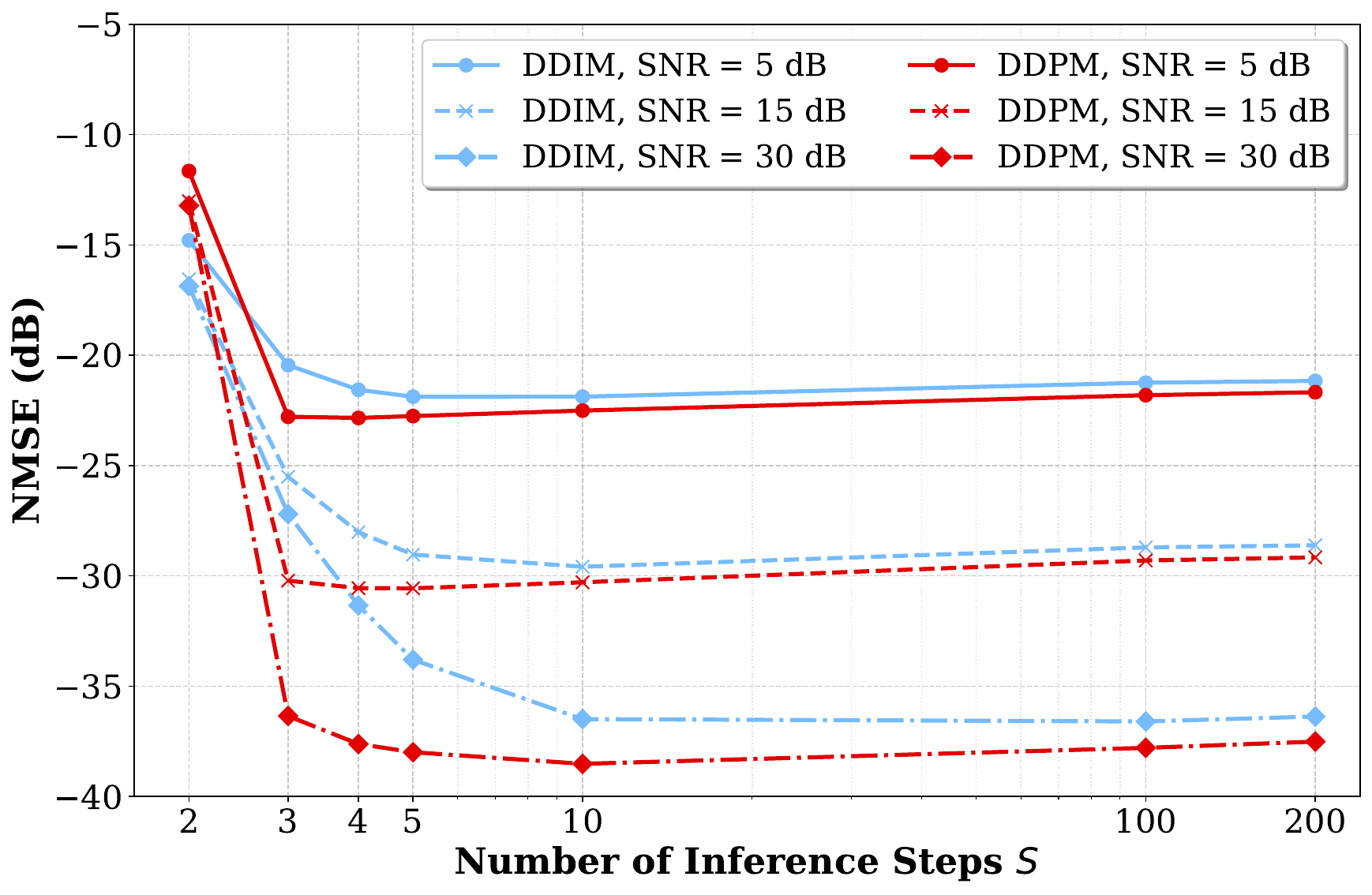}%
  }%
  \hfill 
  \subcaptionbox{$P=16$\label{fig:exp:nmse_vs_steps_16_ddim}}[0.32\linewidth]{%
    \centering
    \includegraphics[width=\linewidth]{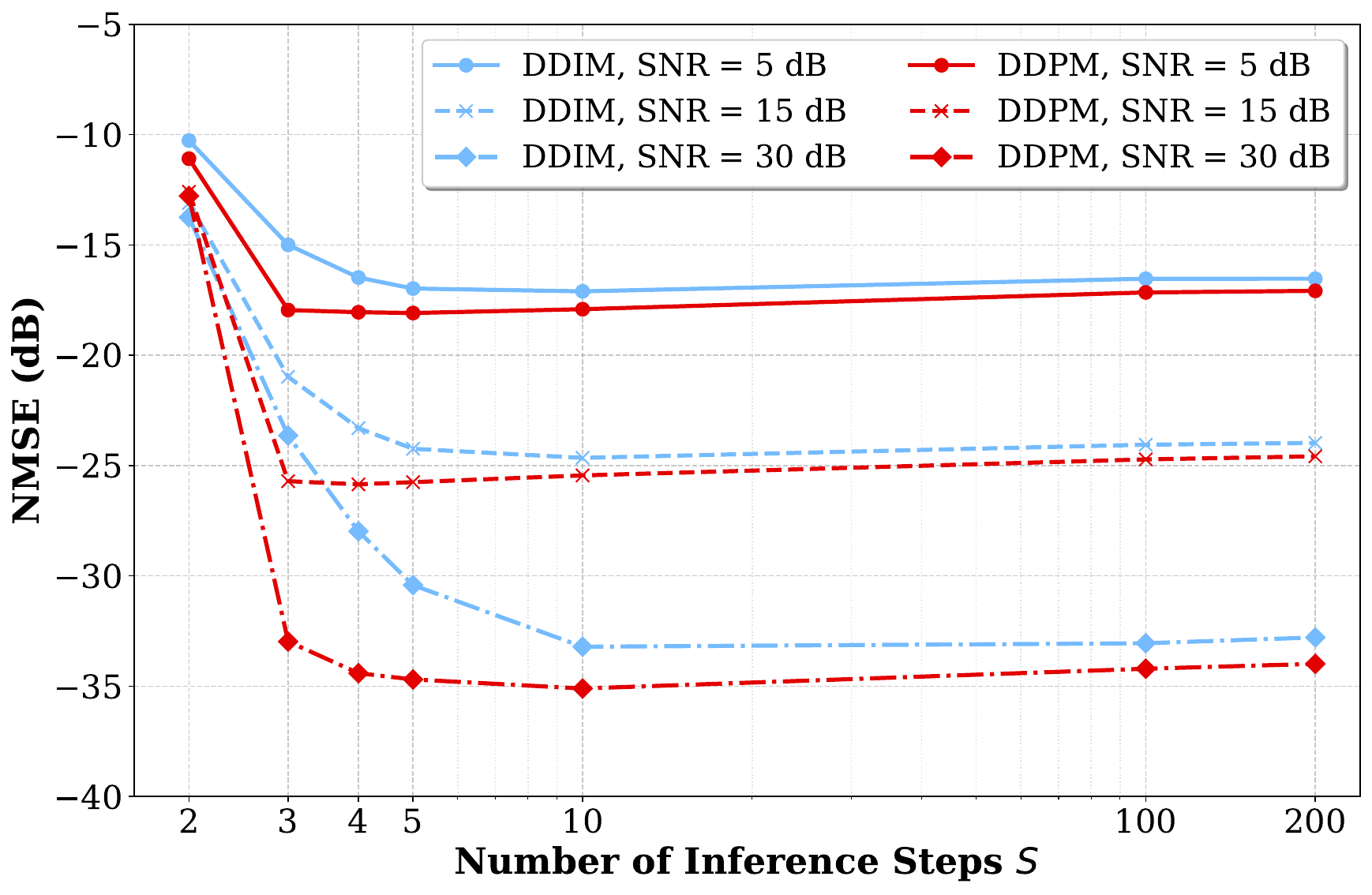}%
  }%
  \hfill 
  \subcaptionbox{$P=32$\label{fig:exp:nmse_vs_steps_32_ddim}}[0.32\linewidth]{%
    \centering
    \includegraphics[width=\linewidth]{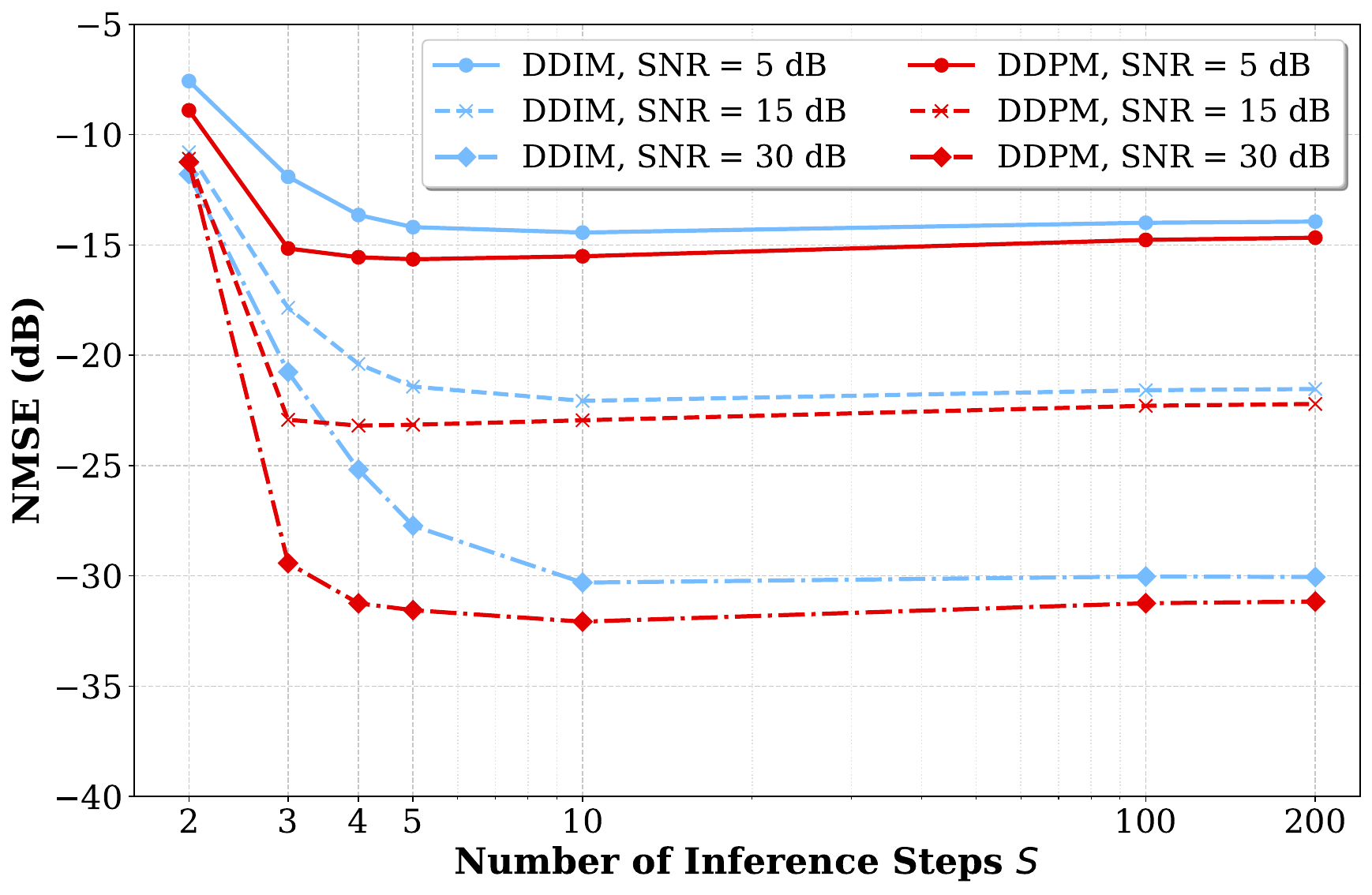}%
  }
  
  \caption{NMSE versus the number of inference steps $S$ of DDIM and that of DDPM under different noise levels.}
  \label{fig:exp:nmse_vs_steps_ddim}
\end{figure*}

\begin{table*}[t!]
    \centering
    \footnotesize
    \renewcommand{\thetable}{\Roman{table}}
    \caption{PERFORMANCE ACROSS DIFFERENT MODEL SIZES, EVALUATED BY NMSE AND $\rho$.}
    \label{exp:tab:model_size}
    \renewcommand{\arraystretch}{1.5} 
    
    \begin{tabularx}{\textwidth}{%
        >{\centering\arraybackslash}p{2.2cm} 
        | *{4}{
            >{\centering\arraybackslash}X 
            >{\centering\arraybackslash}X 
            | 
        } 
        >{\centering\arraybackslash}X 
        >{\centering\arraybackslash}X 
    }
        \hline
        Model
        & \multicolumn{2}{c|}{$K=9$, $d=768$} 
        & \multicolumn{2}{c|}{$K=9$, $d=384$} 
        & \multicolumn{2}{c|}{$K=6$, $d=768$} 
        & \multicolumn{2}{c|}{$K=3$, $d=768$}
        & \multicolumn{2}{c}{$K=3$, $d=384$} \\ 
        \hline
        Number of parameters
        & \multicolumn{2}{c|}{\multirow{2}{*}{136.52 Million}} 
        & \multicolumn{2}{c|}{\multirow{2}{*}{34.49 Million}} 
        & \multicolumn{2}{c|}{\multirow{2}{*}{92.23 Million}} 
        & \multicolumn{2}{c|}{\multirow{2}{*}{47.94 Million}}
        & \multicolumn{2}{c}{\multirow{2}{*}{12.32 Million}}\\ 
        \hline
        Gflops        
        & \multicolumn{2}{c|}{16.55} 
        & \multicolumn{2}{c|}{4.18} 
        & \multicolumn{2}{c|}{11.09} 
        & \multicolumn{2}{c|}{5.63}
        & \multicolumn{2}{c}{1.45}\\ 
        \hline
        Number of inference steps & \multirow{2}{*}{NMSE (dB)} & \multirow{2}{*}{$\rho$} & \multirow{2}{*}{NMSE (dB)} & \multirow{2}{*}{$\rho$} & \multirow{2}{*}{NMSE (dB)} & \multirow{2}{*}{$\rho$} & \multirow{2}{*}{NMSE (dB)} & \multirow{2}{*}{$\rho$} & \multirow{2}{*}{NMSE (dB)} & \multirow{2}{*}{$\rho$} \\
        \hline
        1000 & -33.64 & 0.99975    & -32.24 & 0.99968    & -32.61 & 0.99971    & -31.00 & 0.99955 & -24.59 & 0.99862 \\ 
        10   & -35.06 & 0.99983    & -33.57 & 0.99977    & -33.83 & 0.99978    & -31.93 & 0.99965 & -25.58 & 0.99892 \\ 
        5    & -34.69 & 0.99981    & -33.10 & 0.99975    & -33.42 & 0.99977    & -31.28 & 0.99961 & -25.31 & 0.99886 \\ 
        4    & -34.41 & 0.99980    & -32.58 & 0.99973    & -32.77 & 0.99975    & -30.74 & 0.99956 & -24.54 & 0.99869 \\ 
        3    & -32.98 & 0.99975    & -30.02 & 0.99964    & -29.60 & 0.99969    & -28.88 & 0.99936 & -22.49 & 0.99823 \\ 
        \hline
    \end{tabularx}
\end{table*}

As illustrated in Fig. \ref{fig:exp:nmse_vs_steps_ddpm}, we observe that under different pilot intervals, the NMSE performance of the CDiT exhibits a consistent trend across different inference steps. When the selected steps are 4 or more, the NMSE performance across different inference steps are almost the same. Under the high SNR conditions, noticeable performance degradation begins to occur only at 3 steps. Whereas at low SNR, the degradation caused by very few inference steps is less pronounced. We attribute this to the highly structured properties of channel data, which are distinct from those of images. Such properties enable CDiT to learn the channel's distribution more easily once conditioning is introduced during training. We can also find an important trend that the value of $S$ to obtain the optimal result gradually decreases as the SNR decreases. For example, as the SNR decreases from 30dB to 5dB, the optimal $S$ is reduced from 10 to 5. We believe one of the reasons for the results is the randomness of the noisy estimates. We also have tried other inference time step spacing method such as 'leading' which is widely used in DDIM \cite{Lin_2024_WACV}. We find that the performance of the 'leading' method and 'linspace' is comparable when a larger $S$ is selected. However, a noticeable performance degradation occurs with the 'leading' method when using only 5 steps. Therefore, we consistently employed the 'linspace' method for inference time step spacing in the DDPM inference.

In Figure \ref{fig:exp:nmse_vs_steps_ddim}, we show the impact of value of $S$ for DDIM. The maximum $S$ in the experiment is 200 and we compare the performance of DDIM with that of DDPM when SNR takes the values 5, 15 and 30 dB. The results indicate that the performance of DDIM is worse compared to that of DDPM, particularly under conditions of large pilot intervals and low SNR, where the performance gap widens. Additionally, when the inference steps are less than 10, DDIM exhibits significant performance degradation as number of steps decreases. These findings suggest that in noisy channel estimation tasks, the stochastic noise term in (\ref{eq:DDIM}) plays a critical role in enhancing both the quality and efficiency of generation results. The randomness introduced by noise implies that fixed procedures like DDIM are suboptimal. For DDIM, we use the 'leading' method as it yields better performance.

\subsubsection{Performance versus the patch size and model size}

\begin{figure}[t!]
    \centering
    \includegraphics[width=0.8\linewidth]{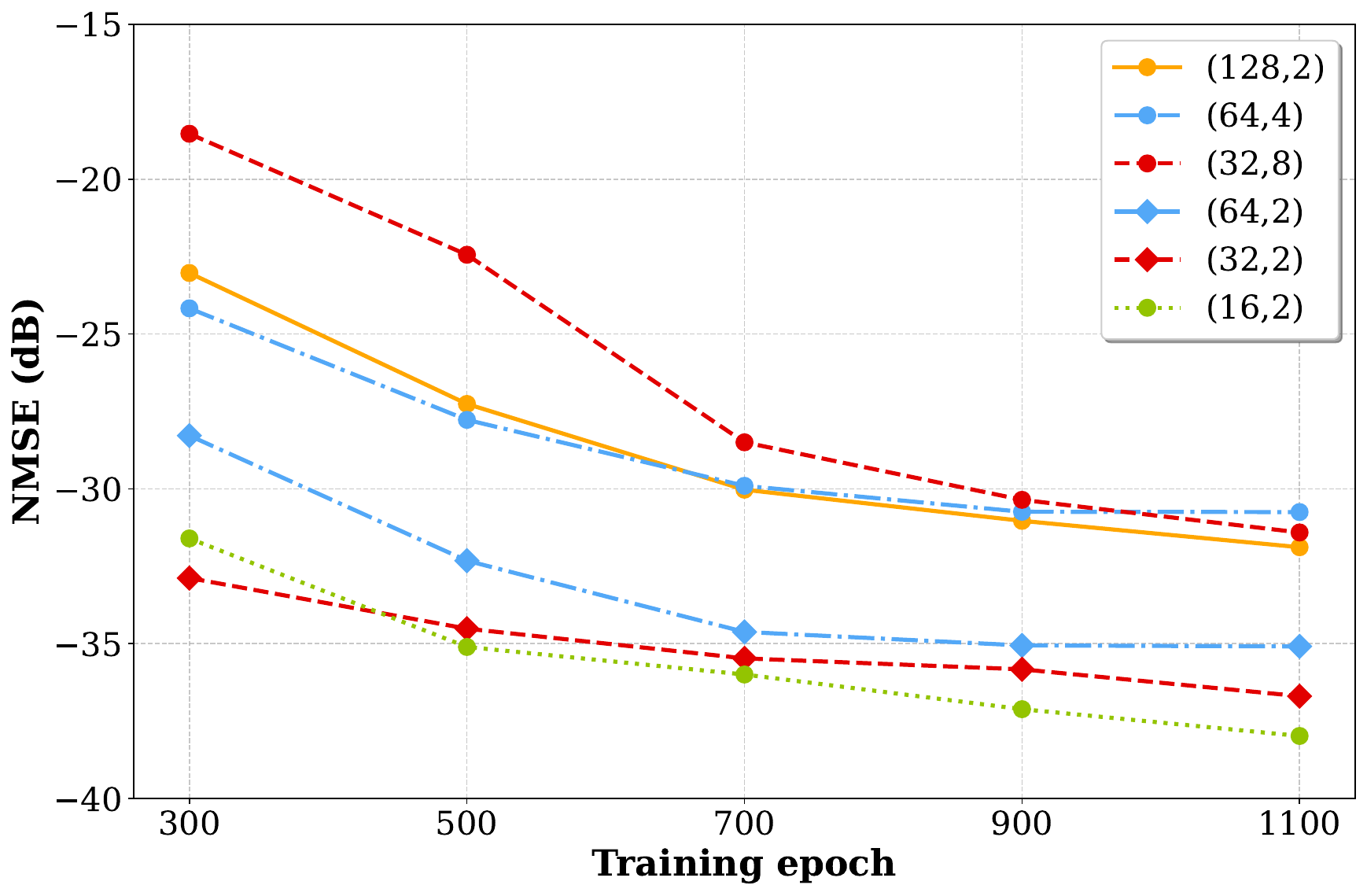}
    \caption{Performance of the CDiT with different training epoch and patch size when $\mathrm{SNR}=30$ dB, $P=16$ and $S = 10$. The model size is fixed which follows the Table \ref{tab:training_params}.}
    \label{fig:exp:trainig epoch and patch size}
    \vspace{-0.3cm}
\end{figure}

 Fig. \ref{fig:exp:trainig epoch and patch size} shows the performance of the CDiT when we change the patch size. We find that reducing the patch size significantly enhances the performance of CDiT. Our experiments reveal that as long as the number of elements within a patch remains consistent, CDiT have the similar performance trained with enough epochs. For example, the performance is almost the same when $(p_\mathrm{f}, p_\mathrm{r})$ takes the values $(128, 2)$, $(64, 4)$ and $(32, 8)$ after 900 training epochs. We also find that when $(p_\mathrm{f}, p_\mathrm{r})$ takes the values $(128, 4)$ and $(64, 8)$, the network will not work satisfactorily, which means that we should not set too large patch size. Different patch size has a minimal impact on model parameters, with fluctuations confined to approximately 1 Million parameters. In contrast, patch size profoundly affects computational costs (Gflops). Smaller patches lead to longer sequence lengths in the attention and substantially higher Gflops and ultimately better performance. Such result is similar to that in DiT \cite{Peebles_2023_ICCV}. We choose $(p_\mathrm{f}, p_\mathrm{r})=(64, 2)$ because there is no significant difference in performance compared to the smaller patch size when the network is trained for sufficient epochs. However, the smaller ones will result in a more pronounced increase in computational cost. 

 Furthermore, as shown in Table \ref{exp:tab:model_size}, we summarize the impact of model size on the performance. We set $(p_\mathrm{f}, p_\mathrm{r})=(64, 2)$ in the experiments. We observe that appropriately reducing either the number of CDiT blocks $K$ or the feature dimension size $d$ to reduce Gflops does not cause significant performance degradation. The performance degradation is almost negligible at low SNR.  Additionally, performance degradation caused by reducing the number of layers is greater than that caused by reducing feature dimensions, even though models with reduced feature dimensions have larger model sizes and higher Gflops than those with reduced layers.

\subsubsection{Performance versus number of pilot patterns} 
 We train different networks with varying numbers of pilot patterns and then compare their performance under a pilot interval of 16. Here, Pilot Patterns A, B, C, D, E, and F correspond to pilot interval sets of $\{2, 3, 4, ..., 64\}$, $\{2, 4, 8, 16, 32, 64\}$, $\{2, 4, 8, 16, 32\}$, $\{4, 8, 16\}$, $\{16\}$, and $\{16\}$, respectively. The difference between E and F is that the starting positions of the pilots are fixed for F, meaning only one pilot pattern exists, whereas those for E are random. According to the training algorithm, if the number of pilot patterns increases while the total training epochs remain fixed, the average number of training iterations per pilot pattern may decrease, potentially leading to inadequate training. Since Pilot Pattern A contains significantly more pilot patterns than the other types, we also conduct an additional experiment with 1500 training epochs for more sufficient training of the network. All other models are trained for 900 epochs. 

\begin{figure}[t!]
    \centering
    
    \begin{subfigure}[b]{\linewidth}
        \centering
        \includegraphics[width=0.8\linewidth]{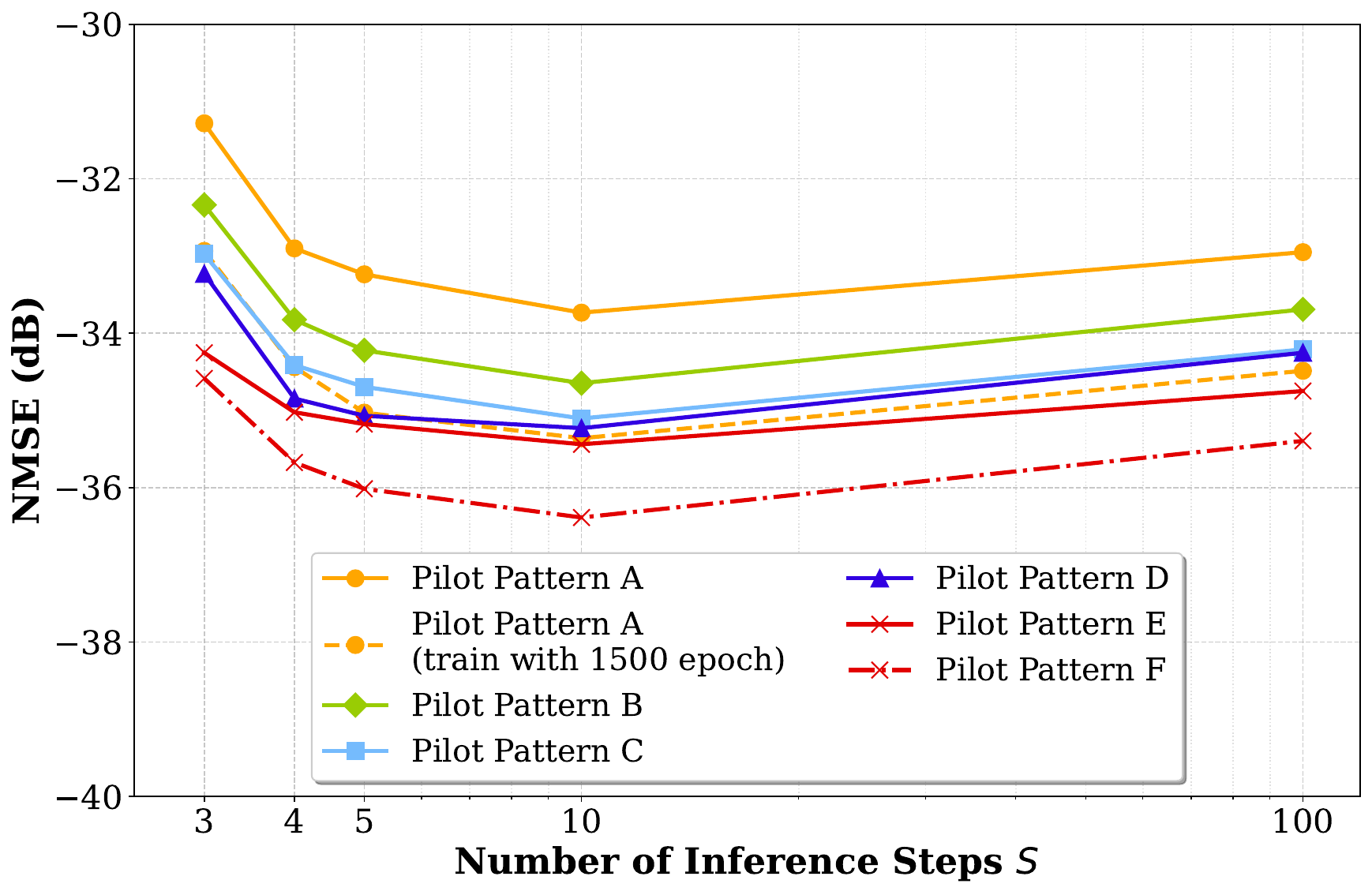}
        \caption{$\mathrm{SNR}=30$ dB}
        \label{fig:exp:mask type} 
    \end{subfigure}
    
    \vspace{0.2cm} 
    
    \begin{subfigure}[b]{\linewidth}
        \centering
        \includegraphics[width=0.8\linewidth]{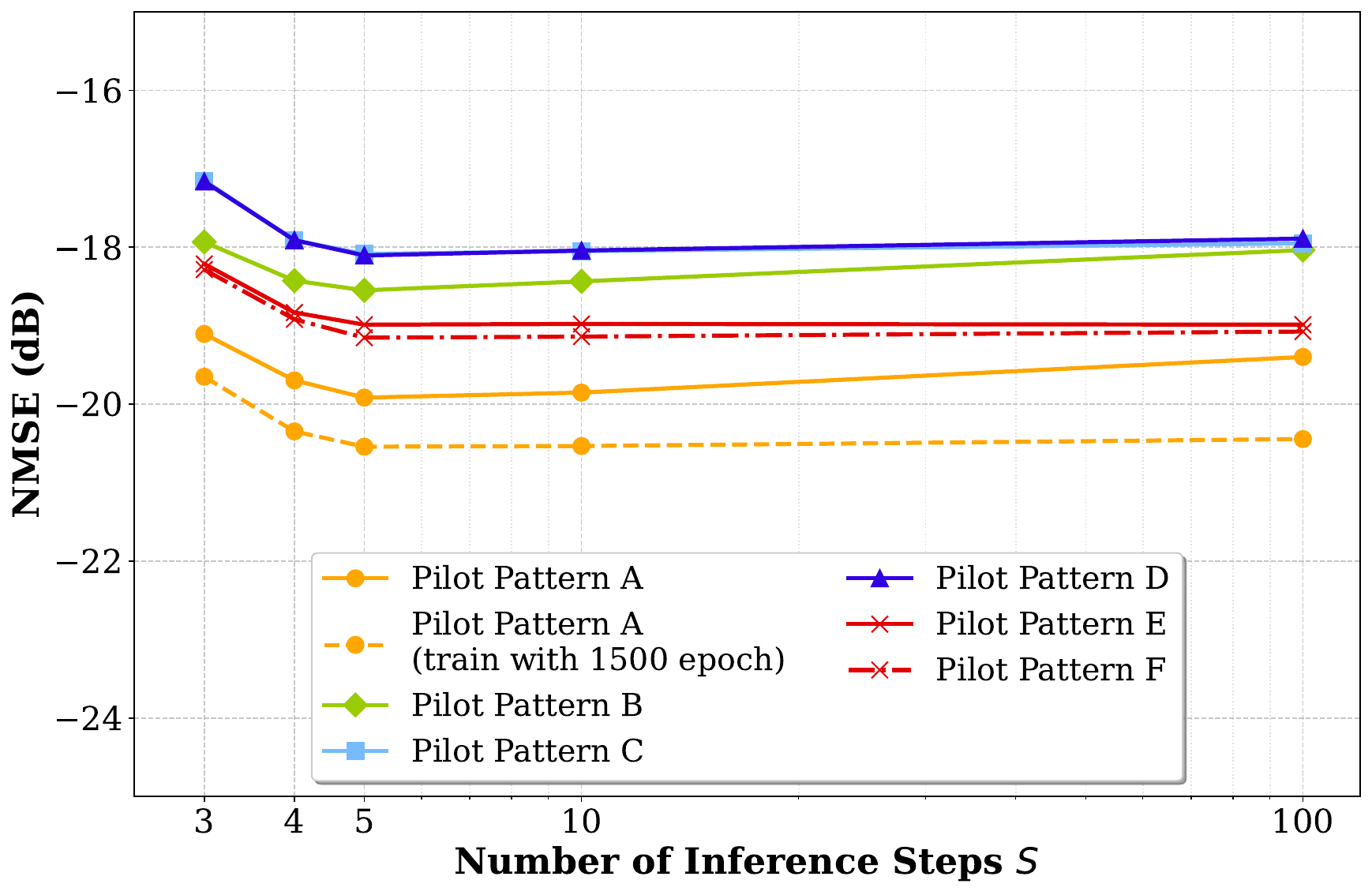}
        \caption{$\mathrm{SNR}=5$ dB}
        \label{fig:exp:mask type low snr} 
    \end{subfigure}
    
    \caption{Performance of the CDiT trained with different number of pilot patterns. The models are tested under the conditions of $P=16$.}
    \label{fig:exp:mask_combined} 
    \vspace{-0.3cm}
\end{figure}

When $\mathrm{SNR}=30$ dB, as shown in Fig. \ref{fig:exp:mask type}, the results indicate that as the number of pilot patterns increases, the model's learning complexity rises, leading to degraded performance under the same number of training epochs. However, in scenarios with a large number of pilot patterns, performance can be effectively improved by increasing the training epochs. In contrast, as illustrated in Fig. \ref{fig:exp:mask type low snr}, when $\mathrm{SNR}=5$ dB, comparing A with other results reveals that a greater variety of pilot patterns actually enhances network performance, even if the average training epochs for a specific pilot pattern are much fewer. Meanwhile, the performance of B, C, and D is inferior to that of E and F. We attribute it to the fact that E and F have received sufficiently training for their specific patterns, where the gain derived from training outweighs the gain derived from the quantity of pilot patterns. The experimental results clearly demonstrate the adaptability of our network to different pilot patterns.


\subsection{Ablation Experiments}
To demonstrate the rationality of the model design and the necessity of the modules, we conducted ablation studies on the relevant modules of CDiT, as indicated by the yellow boxes and numerical labels in Fig. \ref{fig:CDiT model}. Specifically, \ding{182} denotes concatenation and 1×1 conv of weighted mask $\textbf{M}$, \ding{183} denotes noise embedding, \ding{184} denotes class embedding, \ding{185} denotes concatenation of mask $\textbf{M}$, and \ding{186} denotes patchify. The training parameters and datasets in the ablation studies follow the descriptions in Section \ref{exp:experiment detials}. The models are tested at SNR values of 5, 15, and 30 dB. Based on the conclusions from Fig. \ref{fig:exp:nmse_vs_steps_ddpm}, we select the optimal $S$ for each SNR. It is worth noting that in these ablation studies, the different models still maintain the previously observed relationship between the performance and $S$. Specifically, for SNR values of 5 and 15 dB, 
$S$ is set to 5, while for an SNR of 30 dB, that is set to 10. 

\subsubsection{Ablation for the embeddings and concatenation operations}

We first evaluate the impact of modules \ding{182}$\sim$\ding{185} on the model's performance. We train four models by removing each of these four modules independently, and another model by removing all of them. The results are shown in the Table \ref{exp:tab:ablation element}. Here, the first four columns under 'Ablation elements' show the NMSE performance of the model after removing the correspondingly numbered modules, and 'All' represents the NMSE of the model after removing all modules. When module \ding{182} is removed, it means that the raw estimate is directly concatenated with mask. When \ding{185} is removed, it means the result from concatenating and $1\times 1$ conv is directly input into the patchify module.

We find that after removing modules, the results of the ablation study vary with different model sizes. Specifically, when $K=9$ and $d=768$ (in Table \ref{exp:tab:ablation element:large model}), due to the large number of model parameters, removing any single module has almost no impact on the model's performance. However, when $K=9$ and $d=384$ (in Table \ref{exp:tab:ablation element:small model}), the number of parameters is significantly reduced. In this case, we observe that removing any single module results in a significant degradation of the model's performance. Among these, modules \ding{182} and \ding{183} are designed to help the model learn the noise variance in the raw estimates. The results show that both modules are useful. The module \ding{183} effectively improves the model's performance at low SNR, and its impact is more substantial than that of module \ding{182}. When all modules are removed, the conditional information consists solely of the time step $t$ and the raw estimates, where the raw estimates are then directly fed into the patchify layer. This results in significant performance degradation; however, this decline is not merely the summation of the performance drops observed in the previous four modules. This is because the raw estimates inherently contain information regarding the pilot patterns and noise levels, while the removed modules are designed to enhance the model's ability to interpret this conditional information.

\begin{table}[t!]
\centering
\caption{ABLATION STUDY ON CDIT. ALL OF THE MODELS ARE TRAINED FOR 900 EPOCHS. } 
\label{exp:tab:ablation element}

\begin{subtable}{\linewidth}
\centering
\footnotesize
\caption{NMSE (dB) performance when $K=9$ and $d=768$. The number of parameters of CDiT is 136.52 Million.} 
\renewcommand{\arraystretch}{1.5}

\begin{tabularx}{\linewidth}{c|>{\centering\arraybackslash}m{1.2cm}| *{5}{>{\centering\arraybackslash}X} }
\hline
\multirowcell{2}{SNR \\ (dB)} &
\multirowcell{2}{CDiT } &

\multicolumn{5}{c}{Ablation elements} \\
\cline{3-7}
& & \ding{182} & \ding{183} & \ding{184} & \ding{185} & All \\
\hline
{30} & {-35.10} & {-34.86} & {\textbf{-35.15}} & {-34.34} & {-34.99} & {-34.60} \\

{15} & {-25.75} & {\textbf{-25.87}} & {-25.85} & {-25.82} & {-25.69} & {-25.41} \\

{5}  & {-18.09} & {-18.44} & {-18.19} & {$\textbf{-18.59}$} & {-18.05} & {-17.62} \\
\hline
\end{tabularx}
\label{exp:tab:ablation element:large model}
\end{subtable}

\vspace{0.5cm}

\begin{subtable}{\linewidth}
\centering
\footnotesize
\caption{NMSE (dB) performance when $K=9$ and $d=384$. The number of parameters of CDiT is 34.49 Million.} 
\renewcommand{\arraystretch}{1.5}

\begin{tabularx}{\linewidth}{c|>{\centering\arraybackslash}m{1.2cm}| *{5}{>{\centering\arraybackslash}X} }
\hline
\multirowcell{2}{SNR \\ (dB)} &
\multirowcell{2}{CDiT } &
\multicolumn{5}{c}{Ablation elements} \\
\cline{3-7}
& & \ding{182} & \ding{183} & \ding{184} & \ding{185} & All \\
\hline
{30} & {$\textbf{-33.52}$} & {-31.65} & {-31.46} & {-31.43} & {-30.94} & {-30.36} \\

{15} & {$\textbf{-24.87}$} & {-24.31} & {-22.64} & {-24.10} & {-22.99} & {-22.32} \\

{5}  & {$\textbf{-17.47}$} & {-16.82} & {-14.68} & {-16.45} & {-15.32} & {-14.35} \\
\hline
\end{tabularx}
\label{exp:tab:ablation element:small model}
\end{subtable}
\end{table}

\begin{table}[t!]
\centering
\footnotesize
\caption{ABLATION STUDY ON PATCHIFY MODULE \ding{186}. ALL OF THE MODELS ARE TRAINED FOR 900 EPOCHS.}
\renewcommand{\arraystretch}{1.5}
\label{exp:tab:patchify}
\begin{tabular*}{\columnwidth}{@{\extracolsep{\fill}}c|cc|cc|cc}
\hline
& \multicolumn{2}{c|}{$K=9, d=768$} & \multicolumn{2}{c|}{$K=9, d=384$} & \multicolumn{2}{c}{$K=3, d=768$} \\
\hline
SNR (dB) & Diff & Same & Diff & Same & Diff & Same \\
\hline

30 &  $\textbf{-35.10}$   &  -34.09    &  $\textbf{-33.52}$    &   -30.44   &  -31.93    &   $\textbf{-33.05}$   \\
\hline
15 &  $\textbf{-25.75}$   &  -25.42    &  $\textbf{-24.87}$    &   -23.13   &  -24.69    &   $\textbf{-24.98}$   \\
\hline
5  &  $\textbf{-18.09}$   &  -17.76    &  $\textbf{-17.47}$    &   -15.59   &  $\textbf{-17.48}$   &   -17.32    \\
\hline
\end{tabular*}
\end{table}

\subsubsection{Ablation for the patchify module}

We evaluate the performance impact of using identical versus different parameters for the two patchify modules in module \ding{186}; that is, whether to use a single shared patchify module or two distinct ones for feature extraction. If we use identical parameters, the input tensors for the patchify module must have the same dimensions. To align the channel dimension $C$ in this ablation study, we first pass the "Noised Channel" through a $1\times1$ 2D convolutional network, changing its channel count from 2 to 3. As a result, the input dimensions for both patchify operations become ${3\times N_\mathrm{f}\times N_\mathrm{r}}$. The experimental results are shown in Table \ref{exp:tab:patchify}. In the table, 'Diff' refers to using two distinct patchify modules, while 'Same' refers to using a single shared module. 

Similarly, when the model parameters are large, using the same or different patchify modules does not produce a significant difference. However, when the model size is reduced, a significant difference emerges. The feature dimension $d$ essentially determines the representation precision and the model’s capacity for feature decoupling. As $d$ decreases, the representation space becomes more restricted; in this scenario, employing independent patchify modules effectively enforces a structural separation between conditional information and the noisy channel. This prevents feature interference and yields significant performance gains. 
The number of layers $K$ governs the logical learning and non-linear transformation capacity, particularly concerning the depth of information fusion within the cross-attention mechanism. We observed that as $K$ decreases, the advantage of independent patchify modules diminishes. This suggests that without sufficient logical depth to digest and correlate the two feature streams, the benefits of front-end decoupling cannot be fully translated into generative quality.
$d$ and $K$ represent the dual scalability axes of the architecture. $d$ focuses on horizontal information expansion and decoupling at the input stage, while $K$ dictates the vertical depth of logical reasoning. Utilizing independent patchify modules provides a superior representation starting point for horizontal scaling (wider $d$) and provides high-quality, decoupled features that facilitate more efficient vertical scaling (deeper $K$). This design ensures that CDiT maintains robust performance and high information-processing efficiency across various model scales.

\section{Conclusion and Future Directions}
In this paper, we propose CDiT to solve the MIMO-OFDM channel estimation task. Unlike existing approaches that modify the inference process without changing the pre-trained model, we directly incorporate conditional information through a cross-attention module and the embedding method into the model's training. We observe that this approach enables the model to generate optimal results within 10 inference steps, significantly reducing inference overhead. Furthermore, experimental results demonstrate that a single model can handle varying pilot patterns and noise levels, validating the effectiveness and robustness of the proposed method. Within a certain range, the model size can be reduced to balance computational cost and inference latency without significant performance degradation.

Given the powerful capability of DMs in fusing multi-modal information, future work may further explore methods to integrate auxiliary data, such as the positions of UEs and BSs and environmental scenarios. This will enhance channel estimation performance while improving the model's cross-scenario generalization \cite{chen2025analogical}. Also, as DMs are usually time-consuming, techniques such as model distillation, quantization, and few-step generation can further reduce inference time and computational overhead. 


\printbibliography
\end{document}